\pdfoutput=1

\documentclass[letterpaper, 10 pt, conference]{ieeeconf}

\usepackage{graphicx}
\usepackage{booktabs}
\usepackage{adjustbox}
\usepackage{wrapfig}
\usepackage[labelfont=bf,figurename=Fig.,font=small]{caption}
\usepackage{subcaption}
\usepackage{lipsum}
\usepackage{amssymb}
\usepackage{amsmath}
\usepackage{siunitx}
\usepackage{mathtools}
\usepackage{xcolor}
\usepackage{hyperref}
\let\labelindent\relax
\usepackage[inline]{enumitem}
\usepackage[ruled,vlined,linesnumbered]{algorithm2e}
\usepackage[noend]{algpseudocode}
\usepackage{ifthen}
\usepackage{dsfont}
\usepackage[backend=biber,style=numeric-comp,sorting=none,maxbibnames=99]{biblatex}
\usepackage{balance}
\usepackage{dirtytalk}
\usepackage{afterpage}

\usepackage{times}

\allowdisplaybreaks

\newcommand{\E}{\mathbb{E}}

\newcommand{\N}{\mathbb{N}}

\newcommand{\Prob}{\mathbb{P}}

\newcommand{\ra}{\rightarrow}
\newcommand{\R}{\mathbb{R}}
\newcommand{\Ra}{\Rightarrow}

\newtheorem{assumption}{Assumption}

\newtheorem{definition}{Definition}
\newtheorem{lemma}{Lemma}
\newtheorem{theorem}{Theorem}

\newcommand{\graphOrig}{G_O}
\newcommand{\graphMod}{G}
\newcommand{\nodesOrig}{I_O}
\newcommand{\nodesMod}{I}
\newcommand{\arcsOrig}{A_O}
\newcommand{\arcsMod}{A}

\newcommand{\arcsLoadConstraintSet}{\mathcal{W}}
\newcommand{\arcLoadMod}{w}
\newcommand{\arcLoadOrig}{w}
\newcommand{\arcLoadModDiscrete}{W}
\newcommand{\latency}{s}
\newcommand{\toll}{p}
\newcommand{\tollDiscrete}{P}

\newcommand{\costToGo}{z}
\newcommand{\nodeCostToGo}{\varphi}
\newcommand{\routeSetMod}{\textbf{R}}
\newcommand{\routes}{\routeSetMod}

\newcommand{\depth}{\ell}

\newcommand{\height}{m}

\newcommand{\probDist}{\xi}

\newcommand{\nodeLoadIn}{g}

\usepackage{todonotes}
\usepackage{bbm}

\bibliography{references}

\IEEEoverridecommandlockouts
\title{\LARGE \bf Dynamic Tolling in Arc-based Traffic Assignment Models
}

\author{Chih-Yuan Chiu$^{1}$, Chinmay Maheshwari$^{1}$, Pan-Yang Su$^{1}$, and Shankar Sastry$^{1}$
\thanks{Supported by NSF Grant 2031899, Collaborative Research: Transferable, Hierarchical, Expressive, Optimal, Robust, Interpretable Networks.
}
\thanks{$^{1}$Department of Electrical Engineering and Computer Sciences, University of California, Berkeley, CA 94720 (emails: \texttt{\{chihyuan\_chiu, chinmay\_maheshwari, pan\_yang\_su, sastry\} at berkeley dot edu}).}%
}

\begin{document}

\maketitle

\thispagestyle{empty}
\pagestyle{empty}

% \thispagestyle{plain}
% \pagestyle{plain}

% ~\\ \frank{Restore to pagestyle empty during final compile before submitting.} ~\\

%%%%%%%%%%%%%%%%%%%%%%%%%%%%%%%%%%%%%%%%%%%%%%%%%%%%%%%%%%%%%%%%%%%%%%%%%%%%%%%%

\begin{abstract}
Tolling in traffic networks offers a popular measure to minimize overall congestion. Existing toll designs primarily focus on congestion in \emph{route-based} traffic assignment models (TAMs), in which travelers make a single route selection from source to destination. However, these models do not reflect real-world traveler decisions because they preclude deviations from a chosen route, and because the enumeration of all routes is computationally expensive. To address these limitations, our work focuses on \emph{arc-based} TAMs, in which travelers sequentially select individual arcs (or edges) on the network to reach their destination. We first demonstrate that \emph{marginal pricing}, a tolling scheme commonly used in route-based TAMs, also achieves socially optimal congestion levels in our arc-based formulation. Then, we use \textit{perturbed best response} dynamics to model the evolution of travelers' arc selection preferences over time, and a marginal pricing scheme to capture the social planner's adaptive toll updates in response. We prove that our adaptive learning and marginal pricing dynamics converge to a neighborhood of the socially optimal loads and tolls. We then present empirical results that verify our theoretical claims.
\end{abstract}

\section{INTRODUCTION}
\label{sec: Introduction and Related Works}

Mitigating congestion on transportation networks is a key concern in urban planning, since the selfish behavior of individual drivers often significantly increases driving time and pollution levels. Congestion pricing (tolling) is an increasingly popular tool for regulating traffic flows (\cite{manville2021and, nycCongestion}). 
The design of tolls that can effectually induce socially optimal traffic loads requires a realistic traffic assignment model (TAM) that captures travelers' routing preferences. 
% To motivate travelers to adjust their link selection preferences,  To distribute traffic flow in a socially optimal manner, we must then design an adaptive tolling scheme that accounts for an accurate dynamics model of the travelers' preferences, as given by the TAM.

% \frank{To shorten:} 
The classical literature on congestion pricing \cite{Yang1998PrincipleOfMarginalCostPricing,Florian2003NetworkEquilibriumAndPricing, Roughgarden2010AlgorithmicGameTheory} often considers \textit{route-based TAMs}, in which travelers make a single route selection at the origin node of the network, and do not deviate from their selected route until they reach the destination node. However, route-based modeling often requires enumerating all routes in a network, which may be computationally impractical, and do not capture correlations between the total costs of routes that share arcs. To address these issues, this work uses an \textit{arc-based TAM} \cite{Akamatsu1997DecompositionOfPathChoiceEntropy,BaillonCominetti2008MarkovianTrafficEquilibrium,oyama2017discounted,oyama2019prism,Chiu2023ArcbasedTrafficAssignment, kanekoaoptimal} to capture travelers' routing decisions. In this framework, travelers navigate through a traffic network by sequentially selecting among outgoing edges at each intermediate node. 
Designing tolls for arc-based TAMs is relatively under-studied, with the only exception of \cite{kanekoaoptimal} where the authors show that, similar to route based TAMs, marginal tolling also achieves social optimality in arc-based TAMs.

The basic philosophy of toll design is to steer the equilibrium behavior of agents towards social optimality by adding external incentives to their utility functions. However, a key assumption in this setting is that agents always adopt the equilibrium behavior, regardless of the incentives applied. This is not realistic, as real-world agents typically update their strategies from their initial strategies based on repeated interactions, only eventually converging to an equilibrium outcome \cite{fudenberg1998theory}. While there exist learning rules for route-based TAMs which provably converge to the equilibrium strategies \cite{MaheshwariKulkarni2022DynamicTollingforInducingSociallyOptimalTrafficLoads, Sandholm2010PopulationGamesAndEvolutionaryDynamics}, the development of analogous learning mechanisms for arc-based TAMs is relatively recent, e.g., in \cite{Chiu2023ArcbasedTrafficAssignment}, which introduces a perturbed best response based dynamics. Consequently, it is necessary to study tolling in the presence of such dynamic adaptation rules by travelers. 

% Thus, one needs to also tune incentives as travelers are engaging in such adaptive behavior. 

% In particular, \cite{Chiu2023ArcbasedTrafficAssignment} assumes that travelers perform update their strategies via perturbed-best response dynamics, based on the most recently realized costs on each edge of the network, and prove that these dynamics converge to an equilibrium traffic allocation. 

Many prior works design tolls in dynamic environments by using reinforcement learning to iteratively update the toll on each arc. Chen et al. formulated the toll design problem as a Markov Decision Process (MDP) with high-dimensional state and action spaces, and apply a novel policy gradient algorithm to dynamically design tolls \cite{chen2018dyetc}. Mirzaei et al. used policy gradient methods to design incremental tolls on each link based on the difference between the observed and free-flow travel times \cite{Mirzaei2018EnhancedDeltaTolling}. Qiu et al. cast dynamic tolling into the framework of cooperative multi-agent reinforcement learning, and then applies graph convolutional networks to tractably solve the problem \cite{Qiu2019DynamicElectronicTollCollection}. Likewise, Wang et al. use a cooperative actor-critic algorithm to tractably update a dynamic tolling scheme \cite{Wang2022CTRLCooperativeTrafficTollingViaRL}. However, these methods operate on high-dimensional spaces, and are thus often computationally expensive. Moreover, they typically lack theoretical guarantees of convergence. The work most closely related to ours is \cite{MaheshwariKulkarni2022DynamicTollingforInducingSociallyOptimalTrafficLoads} which studies dynamic tolling on parallel-link networks. 

In this work, we study tolling in the arc-based TAM detailed in \cite{Chiu2023ArcbasedTrafficAssignment}. We show that there exists a unique toll that induces socially optimal congestion levels. Furthermore, we propose an adaptive tolling dynamics that steers the travelers' routing preferences towards socially optimal congestion levels on the network. Specifically, we  implement \textit{marginal cost tolling}, via a discrete-time dynamic tolling scheme that adjusts tolls on arcs, with the following key features:
\begin{enumerate}
    \item Tolls are adjusted at each time step towards the direction of the current marginal cost of travel latency.

    \item Tolls are updated at a much slower rate compared to the rate at which travelers update arc selections at each non-destination node (timescale separation).

    \item The toll update of each arc only depends on \say{local information} (in particular, the flow on each arc), and does not require the traffic authority to access the demands of travelers elsewhere on the network.
\end{enumerate}
This form of adaptive tolling was first introduced in \cite{MaheshwariKulkarni2022DynamicTollingforInducingSociallyOptimalTrafficLoads} to study dynamic tolling scheme for parallel-link networks. This work extends the scope of that tolling scheme to bidirectional traffic networks, in the context of arc-based TAMs.

We show that the tolling dynamics converges to a neighborhood of a fixed toll vector, the corresponding equilibrium flows of which we prove to be socially optimal. We also show that the travelers' arc selections converge to a neighborhood of this socially optimal equilibrium flow. Our proof is based on the constant step-size two-timescale stochastic approximation theory \cite{Borkar2009CooperativeDynamics}, which allows us to decouple the toll and arc selection dynamics, and establish their convergence via two separate Lyapunov-based proofs. Although marginal tolling provably leads to socially efficient traffic allocation in a route-based TAM framework \cite{Roughgarden2010AlgorithmicGameTheory}, to the best of our knowledge, this work presents the first marginal tolling scheme that induces socially optimal traffic flows in an arc-based setting.

The rest of the paper is outlined as follows: In Section \ref{sec: Preliminaries} we present the transportation network model we consider in this work and summarize the required preliminaries from \cite{Chiu2023ArcbasedTrafficAssignment} on arc-based TAM. Furthermore, we also introduce the equilibrium concept we consider in this work, along with the notion of social optimality. In Section \ref{sec: Optimal Toll: Existence and Uniqueness}, we present properties of the optimal tolls which induce social optimality in this setup. In Section \ref{sec: Dynamics, Convergence}, we introduce the tolling dynamics and present the convergence results. In Section \ref{sec: Results}, we present a numerical study which corroborate the theoretical findings of this paper. Finally we conclude this paper in \ref{sec: Conclusion and Future Work} and present some directions of future research. 
 % \textbf{2. VCG:}
% Another popular approach for incentive design is to compute the Vickrey Clark Grove payment mechanism \cite{vickrey1961counterspeculation}

\paragraph*{Notation} For each positive integer $n \in \N$, we denote $[n] := \{1, \cdots, n\}$. For each $i \in [n]$ in an Euclidean space $\R^n$, we denote by $e_i$ the $i$-th standard unit vector. Finally, let $\textbf{1}\{\cdot\}$ denote the indicator function, which returns 1 if the input is true and 0 otherwise.

\section{SETUP}
\label{sec: Preliminaries}

% \subsection{Setup} 
% \label{subsec: Setup}

% The following description, of the class of traffic networks considered in this work, is identical to the setup presented in \cite{Chiu2023ArcbasedTrafficAssignment}. We present it for completeness.

\begin{figure}
    \centering
    \includegraphics[scale=0.4]{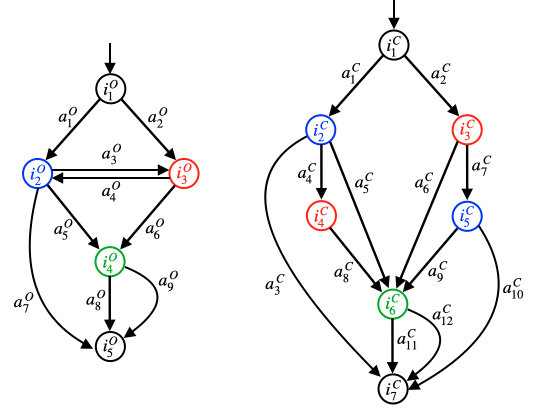}
    \caption{Example of a single-origin single-destination original network $\graphOrig$ (top left, with superscript $O$), and its corresponding condensed DAG, or CoDAG, representation $\graphMod$ (top right, with superscript $C$). Arc correspondences between the two networks are given by Table \ref{table: Equivalent Arcs}, while node correspondences are indicated by color.}
    \label{fig:Front_Figure___Equivalent_DAG}
\end{figure}

Consider a traffic network described by a directed graph $\graphOrig = (\nodesOrig, \arcsOrig)$, where $\nodesOrig$ and $\arcsOrig$ denote nodes and arcs, respectively. An example is shown in Figure \ref{fig:Front_Figure___Equivalent_DAG} (top left); note that $\graphOrig$ can contain bidirectional arcs.
% For every tuple $(o,d)\in \nodesOrig\times \nodesOrig$, let $\nodeLoadIn_o^d$ demand of travelers originate at node $o$  in order to reach destination $d$. 
Let the \textit{origin nodes} and \textit{destination nodes} be two disjoint subsets of $\nodesOrig$. To simplify our exposition, we assume that $\nodesOrig$ contains only one origin $o \in \nodesMod$ and one destination $d \in \nodesMod$, although the results presented below straightforwardly extend to the multiple origin-destination-pair scenario. Travelers navigate through the network, from origin $o$ to destination $d$, by sequentially selecting arcs at every intermediate node. This process produces congestion on each arc, which in turn determines travel times. The \textit{cost} on each arc is then obtained by summing the travel time and toll. Specifically, each arc $a \in A_O$ is associated with a toll $\toll_{a} \in \R^{|\arcsOrig|}$, and a positive, strictly increasing \emph{latency function} $\latency_{a}: [0, \infty) \ra [0, \infty)$, which gives travel time as a function of traffic flow. The cost on arc $a \in \arcsOrig$ is then given by:
\begin{align*}
    c_{a}(w_{a}, \toll_{a}) &= \latency_{\tilde a}(\arcLoadMod_{a}) + \toll_{a}.
\end{align*}
% where  :
% \begin{equation}
% \label{eq: affine latency}
%     \latency_{\tilde a}(\arcLoadMod_{\tilde a}) = \theta_{\tilde a, 1} \arcLoadMod_{\tilde{a}} + \theta_{\tilde a, 0},
% \end{equation}
% for some $\theta_{\tilde a, 1}, \theta_{\tilde a, 0} > 0$. 
Finally, let the demand of  (infinitesimal) travelers entering from origin node \(o\) be denoted by $g_o$. 
% travelers travel from origin $o$ to destination $d$.

Note that sequential arc selection on networks with bidirectional arcs can result in a cyclic route. For example, a traveler navigating the left traffic network in Figure \ref{fig:Front_Figure___Equivalent_DAG} using sequential arc selection may cycle between nodes $i_2^O$ and $i_3^O$. To resolve this issue, we consider arc selection on the \textit{condensed DAG (CoDAG) representation} of the original network $\graphOrig$, a directed acyclic graph (DAG) representation, as proposed in \cite{Chiu2023ArcbasedTrafficAssignment}. The Condensed DAG representation preserves all acyclic routes from origin $o$ to destination $d$ in $\graphOrig$, but precludes cyclic routes by design. Details regarding the construction and properties of CoDAG representations are provided in \cite{Chiu2023ArcbasedTrafficAssignment}, Section II.

\vspace{1mm}

\renewcommand{\arraystretch}{1.4}
\begin{table}[h!]
\centering
\caption{Arc correspondences between the graphs in Figure \ref{fig:Front_Figure___Equivalent_DAG}: The original network (top left) and the CoDAG (top right).}
\label{table: Equivalent Arcs}
\resizebox{\columnwidth}{!}{
\begin{tabular}{||c||c|c|c|c|c|c|c|c|c||}
    \hline
    Original & $a_1^O$ & $a_2^O$ & $a_3^O$ & $a_4^O$ & $a_5^O$ & $a_6^O$ & $a_7^O$ & $a_8^O$ & $a_9^O$ \\
    \hline
    CoDAG & $a_1^T$ & $a_2^T$ & $a_4^T$ & $a_7^T$ & $a_5^T$ & $a_6^T$ & $a_3^T$ & $a_{11}^T$ & $a_{12}^T$ \\
    & & & & & $a_9^T$ & $a_8^T$ & $a_{10}^T$ & & \\
    \hline
\end{tabular}
}
\end{table}

We define $[\cdot]: \arcsMod \ra \arcsOrig$ to be a map from each CoDAG arc $a \in \arcsMod$ to the corresponding arc in the original graph, $[a] \in \arcsOrig$ (as shown in Table \ref{table: Equivalent Arcs}). For each arc $a \in \arcsMod$, let $i_a$ and $j_a$ denote the start and terminal nodes, and for each node $i \in I$, let $\arcsMod_i^-, \arcsMod_i^+ \subset \arcsMod$ denote the set of incoming and outgoing arcs. 

\subsection{Cost Model}
\label{subsec: Cost Model}

Below, we assume that every traveler has access to $\graphOrig$, and to the same CoDAG representation $\graphMod = (\nodesMod, \arcsMod)$ of $\graphOrig$; in particular, $\graphMod$ is used to perform sequential arc selection to generate acyclic routes. The travelers' aggregative arc selections generate network congestion. Specifically, for each $a\in A$, let the \textit{flow} or \textit{congestion level} on arc $a$ be denoted by $w_a$, and let the total flow on the corresponding arc in the original network be denoted, with a slight abuse of notation, by $w_{[a]} := \sum_{a' \in [a]} w_{a'}$\footnote{Unlike existing TAMs, in our model, the latency of arcs in $G$ can be coupled, since multiple copies of the same arc in $G_O$ may exist in $G$.}. Travelers perceive the cost on each arc $a \in A$ as:
\begin{align*}
    \tilde c_{[a]}(\arcLoadMod_{[a]}, \toll_{[a]}) &:= c_{[a]}(\arcLoadMod_{[a]}, \toll_{[a]}) + \nu_{a} \\
    &= s_{[a]}(w_{[a]}) + \toll_{[a]} + \nu_{a},  
\end{align*}
where $\nu_{a}$ is a zero-mean random variable. 
At each non-destination node $i \in I \backslash \{d\}$, travelers select among outgoing nodes $a \in A_i^+$ by comparing their perceived cost-to-go $\tilde \costToGo_a: \R^{|A|}\times \R^{|A_O|} \ra \R$, given recursively by:     
\begin{alignat}{2} \label{eq: LTGPerturbed}
    \tilde{z}_a(\arcLoadMod, \toll) &:= \tilde{s}_{[a]}(w_{[a]}) + \toll_{[a]} + \min_{a' \in A_{j_a}^+} \tilde{z}_{a'}(\arcLoadMod, \toll), \hspace{3mm} &&j_a \ne d, \\ \nonumber
    \tilde{z}_a(\arcLoadMod, \toll) &:= \tilde{s}_{[a]}(w_{[a]}) + \toll_{[a]}, &&j_a = d.
\end{alignat}
Consequently, the fraction of travelers who arrives at \(i\in I\backslash\{d\}\) and choose arc \(a\in A_i^+\) is given by:
\begin{align}\label{eq: Pij, def}
    P_{ij_a} := \Prob(\tilde z_a \leq \tilde z_{a'}, \hspace{0.5mm} \forall \hspace{0.5mm} a' \in A_i^+).
\end{align}
% and selects the arc that maximizes this probability among all arcs in $a \in A_i^+$. 
An explicit formula for the probabilities $\{P_{ij_a}: a \in A_i^+\}$, in terms of the statistics of $\tilde z_a$,
is provided by the discrete-choice theory 
\cite{BenAkiva1985DiscreteChoiceAnalysis}. In particular, 
define $z_a(w) := \E[\tilde z_a(w)]$ and $\epsilon_a := \tilde z_a(w) - z_a(w)$, 
and define the latency-to-go at each node by:
\begin{align} \label{eq: PhiFunc}
    \nodeCostToGo_{i}(\{\costToGo_{a'}(\arcLoadMod,\toll): a' \in \arcsMod_i^+ \}) = \E\Bigg[ \min_{a' \in \arcsMod_i^+} 
    % \Big\{ 
    \tilde \costToGo_{a'}(\arcLoadMod, \toll) 
    % + \epsilon_{a'} \Big\}
    \Bigg].
\end{align}
Then, from discrete-choice theory \cite{BenAkiva1985DiscreteChoiceAnalysis}:
\begin{align}\label{eq: ProbTransition}
    P_{ij_a}  = \frac{\partial \varphi_i}{\partial z_a}(z), \quad i\in I\backslash\{d\}, a\in A_i^{+},
\end{align}
where, with a slight abuse of notation, we write $\nodeCostToGo_i(\costToGo)$ for $\nodeCostToGo_{i}(\{\costToGo_{a'}: a' \in \arcsMod_i^+ \})$. 

% Consequently, the fraction of travelers who arrive at \(i\in I\backslash\{d\}\) and choose arc \(a\in A_i^+\) is given by:
% \begin{align}\label{eq: Pij, as derivative}
%     P_{ij_a} := \Prob(\tilde z_a \leq \tilde z_{a'}, \hspace{0.5mm} \forall \hspace{0.5mm} a' \in A_i^+).
% \end{align}
% % and selects the arc that maximizes this probability among all arcs in $a \in A_i^+$. 
% An explicit formula for the probabilities $\{P_{ij_a}: a \in A_i^+\}$ in terms of the statistics of $\tilde z_a$,
% is provided by the discrete-choice theory \cite{BenAkiva1985DiscreteChoiceAnalysis}. In particular, define $z_a(w,p) := \E[\tilde z_a(w,p)]$ and $\epsilon_a := \tilde z_a(w, p) - z_a(w, p)$, and define the latency-to-go at each node by:
% \begin{align} \label{eq: PhiFunc}
%     \nodeCostToGo_{i}(\{\costToGo_{a'}(\arcLoadMod,\toll): a' \in \arcsMod_i^+ \}) = \E\Bigg[ \min_{a' \in \arcsMod_i^+} 
%     % \Big\{ 
%     \tilde \costToGo_{a'}(\arcLoadMod, \toll) 
%     % + \epsilon_{a'} \Big\}
%     \Bigg],
% \end{align}
% Then, from discrete-choice theory \cite{BenAkiva1985DiscreteChoiceAnalysis}:
% \begin{align}\label{eq: ProbTransition}
%     P_{ij_a}  = \frac{\partial \varphi_i(z)}{\partial z_a}, \quad i\in I\backslash\{d\}, a\in A_i^{+},
% \end{align}
% where, with a slight abuse of notation, we write $\nodeCostToGo_i(\costToGo)$ for $\nodeCostToGo_{i}(\{\costToGo_{a'}: a' \in \arcsMod_i^+ \})$.

To obtain a closed-form expression of $\varphi$, we employ the \emph{logit Markovian model} \cite{Akamatsu1997DecompositionOfPathChoiceEntropy,BaillonCominetti2008MarkovianTrafficEquilibrium}, under which the noise terms $\epsilon_a$ are described by the Gumbel distribution with scale parameter $\beta$. As a result, the expected minimum cost-to-go $\costToGo_a: \R^{|\arcsMod|} \times \R^{|\arcsMod_O|}\ra \R$,  associated with traveling on each arc $a \in \arcsMod$, assumes the following form:
{\small
\begin{align} \label{Eqn: CostToGo}
    &\hspace{5mm} \costToGo_a(\arcLoadMod, \toll) \\ \nonumber
    &= \latency_{[a]} \Bigg( \sum_{\bar a \in [a]} \arcLoadMod_{\bar a} \Bigg) + \toll_{[a]} - \frac{1}{\beta} \ln\Bigg( \sum_{a' \in \arcsMod_{j_a}^+} e^{-\beta \costToGo_{a'}(\arcLoadMod, \toll)} \Bigg).
    % \tag{CTG} 
\end{align}
}
% \begin{align} \label{Eqn: CostToGo}
%     \costToGo_a(\arcLoadMod, \toll) &= \latency_{[a]} \Bigg( \sum_{\bar a \in [a]} \arcLoadMod_{\bar a} \Bigg) + \toll_{[a]} \\ \nonumber
%     &\hspace{1cm} - \frac{1}{\beta} \ln\Bigg( \sum_{a' \in \arcsMod_{j_a}^+} e^{-\beta \costToGo_{a'}(\arcLoadMod, \toll)} \Bigg).
%     % \tag{CTG} 
% \end{align}
Note that \eqref{Eqn: CostToGo} is well-posed, as $\costToGo_a$ can be recursively computed from the destination back to the origin (\cite{Chiu2023ArcbasedTrafficAssignment}, Section III).
% (More details regarding properties of the height of DAGs are given in \cite{Chiu2023ArcbasedTrafficAssignment}, Appendix A, Propositions 1 and 2.)

\subsection{CoDAG Equilibrium}
\label{subsec: CoDAG Equilibrium}
Here, we define the \emph{condensed DAG (CoDAG) equilibrium} (Definition \ref{Def: CoDAG Equilibrium}), based on the CoDAG representation of the original traffic network.  Specifically, we show that the CoDAG equilibrium exists, is unique, and solves a strictly convex optimization problem (Theorem \ref{Thm: CoDAG is unique minimizer of F}).

\begin{definition}[\textbf{Condensed DAG Equilibrium}] \label{Def: CoDAG Equilibrium}
% Let $G_O$ be the graph representing the transportation network and $G$ be the CoDAG representation.
Fix a toll vector $\toll \in \R^{|\arcsOrig|}$, and fix $\beta > 0$. We call an arc-flow vector $\bar \arcLoadMod^\beta(\toll) \in \R^{|\arcsMod|}$ a \emph{Condensed DAG (CoDAG) equilibrium at $\toll$} if, for each $i \in \nodesMod \backslash \{d\}$, $a \in \arcsMod_i^+$:
\begin{align} \label{Eqn: MTE}
   &\hspace{5mm} \bar \arcLoadMod_a^\beta(\toll) \\
   &= \left(\nodeLoadIn_{i} + \sum_{a' \in \arcsMod_{i}^+} \bar \arcLoadMod_{a'}^\beta(\toll) \right)  \frac{\exp(-\beta \costToGo_a(\bar \arcLoadMod^\beta(\toll), \toll))}{\sum_{a' \in \arcsMod_{i_a}^+} \exp(-\beta \costToGo_{a'}(\bar \arcLoadMod^\beta(\toll), \toll))},
    % \\ \nonumber
    % \arcLoadMod_a &= \nodeLoadIn_{i_a} \cdot \frac{\exp(-\beta \costToGo_a((\arcLoadMod)))}{\sum_{\bar a \in \arcsMod_{i_a}^+} \exp(-\beta \costToGo{\bar a} ((\arcLoadMod)))}, \hspace{5mm} \forall \hspace{0.5mm} a \in \arcsMod_{o}^+.
\end{align}
where $g_i = g_0 \cdot \textbf{1}(i=o)$, and $\arcLoadMod \in \arcsLoadConstraintSet$, where:
\begin{align} \label{Eqn: Def, W}
    \arcsLoadConstraintSet := &\Bigg\{ \arcLoadMod \in \R^{|\arcsMod|}: \sum_{a \in \arcsMod_i^+} \arcLoadMod_a = \sum_{a \in \arcsMod_i^-} \arcLoadMod_a, \hspace{0.5mm} \forall \hspace{0.5mm} i \ne o, d, \\ \nonumber
    &\hspace{5mm} \sum_{a \in \arcsMod_o^+} \arcLoadMod_a = \nodeLoadIn_o, \hspace{1mm} \arcLoadMod_a \geq 0, \hspace{0.5mm} \forall \hspace{0.5mm} a \in \arcsMod \Bigg\}
\end{align}
characterizes the conservation of flow in the CoDAG $\graphMod$. Note that $\arcsLoadConstraintSet$ is convex and compact. 
\end{definition}

\vspace{0.5mm}
At a CoDAG equilibrium $\bar{w}^{\beta}(\toll)$, the fraction of travelers at any intermediate node $i\in I\backslash\{d\}$ who selects an arc $a\in A_i^+$ is given by $\bar \probDist_a^\beta(\toll)$, as defined below:
\begin{align*}
    \bar \probDist_a^\beta(\toll) := \frac{\bar \arcLoadMod_a^\beta(\toll)}{\sum_{a' \in A_{i}^+} \bar \arcLoadMod_{a'}^\beta(\toll)}.
\end{align*}

The CoDAG equilibrium bears some resemblance to the Markovian Traffic Equilibrium (MTE) introduced in Baillon and Cominetti \cite{BaillonCominetti2008MarkovianTrafficEquilibrium}. However, the CoDAG formulation by design precludes the possibility of assigning cyclic routes, and is capable of capturing couplings between arcs in the CoDAG $\graphMod$ that correspond to the same arc in the original network $\graphOrig$ (see  \cite{Chiu2023ArcbasedTrafficAssignment}, Remark 6).

Below, we show that, given any CoDAG representation $\graphMod$ of an original network $\graphOrig$ and any fixed toll vector $\toll \in \R^{|\arcsOrig|}$, the CoDAG equilibrium exists and is unique. Specifically, the CoDAG equilibrium is the unique minimizer of a strictly convex optimization problem over a compact set. This characterization provides powerful insight into the mathematical properties of the CoDAG equilibrium flow, and its dependence on the toll vector. These properties will be used in our work to establish the existence of an optimal toll (Theorem \ref{Thm: Fixed Point, p}) and the convergence of our discrete-time toll dynamics to the optimal toll (Theorem \ref{Thm: Convergence, w, p, discrete}).

For each $[a] \in \arcsOrig$, define $F: \arcsLoadConstraintSet \times \R^{|\arcsOrig|} \ra \R$ by:
{\small
\begin{align} \nonumber
    &F(\arcLoadMod, \toll) \\ \nonumber
    = \hspace{0.5mm} &\sum_{[a]\in \arcsOrig} \int_0^{\arcLoadMod_{[a]}} \big[ \latency_{[a]}(u) + p_{[a]} \big] \hspace{0.5mm} du \\ \label{Eqn: Def, F}
    & + \frac{1}{\beta} \sum_{i \ne d} \Bigg[ \sum_{a \in \arcsMod_i^+} \arcLoadMod_a \ln \arcLoadMod_a - \Bigg( \sum_{a \in \arcsMod_i^+} w_a\Bigg) \ln \Bigg( \sum_{a \in \arcsMod_i^+} w_a \Bigg) \Bigg].
\end{align}
}
% where $\arcLoadMod_{\arcsMod_i^+} \in \R^{|\arcsMod_i^+|}$ denotes the components of $\arcLoadMod$ corresponding to arcs in $\arcsMod_i^+$.

\begin{theorem} \label{Thm: CoDAG is unique minimizer of F}
For each fixed toll vector $\toll \in \R^{|\arcsOrig|}$, the corresponding CoDAG equilibrium $\bar w^\beta(\toll) \in \arcsLoadConstraintSet$ exists, is unique, and is the unique minimizer of $F(\cdot, \toll)$ over $\arcsLoadConstraintSet$.
\end{theorem}

\begin{proof}(\textbf{Proof Sketch})
The proof parallels that of \cite{Chiu2023ArcbasedTrafficAssignment}, Theorem 1 and Lemma 1. For details, please see \cite{Chiu2023ArcbasedTrafficAssignment}, Section III and Appendix B.
\end{proof}

% To prove Theorem \ref{Thm: CoDAG is unique minimizer of F}, we first show that for each fixed toll vector $\toll \in \R^{|\arcsOrig|}$, the map $F(\cdot, \toll)$ is strictly convex over $\arcsLoadConstraintSet$ (Lemma \ref{Prop: Strict Convexity of F}). Therefore, $F$ has a unique minimizer in $\arcsLoadConstraintSet$. It then suffices to show that the CoDAG equilibrium definition (Definition \ref{Def: CoDAG Equilibrium}) matches the Karush-Kuhn-Tucker (KKT) conditions for the optimization problem \eqref{Eqn: Def, F}.  

% % To establish the proof of Theorem \ref{Prop: CoDAG is unique minimizer of F}, a crucial step is to show that $F$ is strictly convex over $\arcsLoadConstraintSet$.

% \begin{lemma} \label{Prop: Strict Convexity of F}
% For any fixed toll vector $\toll \in \R^{|\arcsOrig|}$, the map $F(\cdot, \toll): \arcsLoadConstraintSet \ra \R$ is strictly convex.
% \end{lemma}

% \begin{proof}(\textbf{Proof Sketch})
% The proof parallels that of \cite{Chiu2023ArcbasedTrafficAssignment}, Lemma 1. For details, please see \cite{Chiu2023ArcbasedTrafficAssignment}, Appendix B.
% \end{proof}

\subsection{Social Optimality}
\label{subsec: Social Optimality and Optimal Toll}

% The purpose of the tolling mechanism is to adaptively modify the toll vector $\toll(t) \in \R^{|\arcsOrig|}$ to induce flow allocations that satisfy a notion of social optimality. 
We now describe the socially optimal flow which would lead to the most efficient use of the transportation network. More specifically, we define below the notion of \textit{perturbed social optimality} considered in our work.

\vspace{1mm}
\begin{definition}[\textbf{Perturbed Socially Optimal Flow}] \label{Def: Perturbed Socially Optimal Flow}
We define a \emph{perturbed socially optimal flow} with regularization parameter $\beta > 0$ to be a minimizer of the following convex optimization problem:
{\small
\begin{align*}
    &\min_{w \in \arcsLoadConstraintSet} \hspace{3mm} \sum_{[a] \in \arcsOrig} \arcLoadOrig_{[a]} \cdot \latency_{[a]}(\arcLoadOrig_{[a]}) \\
    & \hspace{5mm} + \frac{1}{\beta} \sum_{i \ne d} \Bigg[ \sum_{a \in A_i^+} \arcLoadMod_a \ln \arcLoadMod_a - \Bigg(\sum_{a \in A_i^+} \arcLoadMod_a \Bigg) \ln \Bigg(\sum_{a \in A_i^+} \arcLoadMod_a \Bigg) \Bigg],
\end{align*}}
with $\arcsLoadConstraintSet$ given by \eqref{Eqn: Def, W}, and $\arcLoadOrig_{[a]} := \sum_{a' \in [a]} \arcLoadMod_{a'}$, as defined above.
\end{definition}

In words, perturbed social optimality is characterized as the total latency experienced by travelers on each arc of the CoDAG $\graphMod$, augmented by an entropy term with regularization parameter $\beta$ which captures stochasticity in the travelers' arc selections.

\section{OPTIMAL TOLL: EXISTENCE AND UNIQUENESS}
\label{sec: Optimal Toll: Existence and Uniqueness}

Below, we characterize the \textit{optimal toll} $\bar \toll \in \R^{|\arcsOrig|}$ for which the corresponding CoDAG equilibrium $\bar \arcLoadMod^\beta(\bar \toll)$ is perturbed socially optimal (see Definition \ref{Def: Perturbed Socially Optimal Flow}).
% In particular, we show that there exists a unique toll vector $\tilde \toll \in \R^{|A_0|}$ whose corresponding CoDAG flow values $\tilde \arcLoadMod^\beta(\tilde \toll) \in \R^{|\arcsMod|}$ are perturbed socially optimal. 
Throughout the rest of the paper, we call $\bar \toll$ the \textit{optimal toll}.

\begin{theorem} \label{Thm: Fixed Point, p}
There exists a unique toll vector $\bar \toll \in \R^{|A_0|}$ that satisfies the following fixed-point equation:
\begin{align} \label{Eqn: Toll, Fixed Point Equation}
    \bar \toll_{[a]} = \bar w^{\beta}_{[a]}(\bar \toll) \cdot \frac{d\latency_{[a]}}{d w} \bar w^{\beta}_{[a]}(\bar \toll), \hspace{5mm} \forall \hspace{0.5mm} a \in \arcsMod.
\end{align}
Moreover, $\bar \arcLoadMod^\beta(\bar \toll)$, the CoDAG equilibrium flow distribution corresponding to $\bar \toll$, is the perturbed socially optimal flow with regularization $\beta$.
\end{theorem}

% \begin{proof}
    
% \end{proof}

To prove Theorem \ref{Thm: Fixed Point, p}, we first show that $\bar \arcLoadMod^\beta(\toll)$ is continuous and monotonic in the toll $\toll$ (Lemmas \ref{Lemma: C1 of CoDAG Equilibrium} and \ref{Lemma: Monotonicity of CoDAG Equilibrium}). Then, we use these properties to establish the existence and uniqueness of a toll vector $\bar \toll \in \R^{|\arcsOrig|}$ satisfying the fixed-point equation \eqref{Eqn: Toll, Fixed Point Equation}  (Lemma \ref{Lemma: p bar uniquely exists}). Finally, we prove that the CoDAG equilibrium flow allocation $\bar \arcLoadMod^\beta(\bar \toll)$ corresponding to $\bar \toll$ is perturbed socially optimal (Lemma \ref{Lemma: p bar is Perturbed Socially Optimal}). 

% Below, we begin by establishing that the CoDAG equilibrium $\bar \arcLoadMod^\beta(\toll)$ is a monotonic function of the toll $\toll \in \R^{|\arcsOrig|}$.

Below, we begin by establishing that the CoDAG equilibrium $\bar \arcLoadMod^\beta(\toll)$ is a continuously differentiable and monotonic function of the toll $\toll \in \R^{|\arcsOrig|}$.

\begin{lemma} \label{Lemma: C1 of CoDAG Equilibrium}
$\bar \arcLoadMod^\beta(\toll)$ is continuously differentiable in $\toll$.
\end{lemma}

\vspace{1mm}
\begin{proof}(\textbf{Proof Sketch})
For each fixed toll vector $\toll \in \R^{|\arcsOrig|}$, the corresponding CoDAG equilibrium $\bar \arcLoadMod^\beta(\toll)$ uniquely solves the KKT conditions of the optimization problem of minimizing $F(\cdot, \toll)$ over $\arcsLoadConstraintSet$ (Theorem \ref{Thm: CoDAG is unique minimizer of F}). We write these KKT conditions as an \textit{implicit function} $J: \R^{|\arcsMod|} \times \R^{|\arcsOrig|} \ra \R^{|\arcsMod|}$ of the flow and tolls $(\arcLoadMod, \toll)$:
\begin{align*}
    J(w, p) = \textbf{0},
\end{align*}
where $\textbf{0}$ denotes the $|\arcsMod|$-dimensional zero vector. We can then derive an explicit expression for $\frac{d \bar \arcLoadMod^\beta}{d\toll}(\toll)$ at each $\toll \in \R^{|\arcsOrig|}$ by proving that:
\begin{align*}
    \frac{\partial J}{\partial w}\big( \bar \arcLoadMod^\beta(p), p \big) \in \R^{|\arcsMod| \times |\arcsMod|}
\end{align*}
is non-singular for each fixed $\toll$, and invoking the Implicit Function Theorem. For details, please see Appendix \ref{subsubsec: A1, Proof of Lemma: C1 of CoDAG Equilibrium} \cite{Chiu2023DynamicTollingInArcBasedTAMs}.
\end{proof}

% \begin{remark}
% To prove that the fixed-point equation (\ref{Eqn: Toll, Fixed Point Equation}) yields a unique solution, it actually suffices to show that the CoDAG equilibrium $\bar \arcLoadMod^\beta(\toll)$ is \textit{continuous} in the toll vector $\toll \in \R^{|\arcsOrig|}$. However, the fact that $\frac{d \bar \arcLoadMod^\beta}{d \toll}(\toll) \in \R^{|\arcsOrig| \times |\arcsOrig|}$ is well-defined and continuous at each $\toll$ will be required in later sections to establish that the discrete-time toll dynamics converge to a neighborhood of the optimal toll.
% \end{remark}

\begin{lemma} \label{Lemma: Monotonicity of CoDAG Equilibrium}
For any $\toll, \toll' \in \R^{|A_0|}$:
\begin{align*}
    \sum_{a \in A} \Big( \bar \arcLoadMod_a^\beta(p') - \bar \arcLoadMod_a^\beta(p) \Big) (\toll_{[a]}' - \toll_{[a]}) \leq 0.
\end{align*}
\end{lemma}

\vspace{1mm}
\begin{proof}(\textbf{Proof Sketch})
By Theorem \ref{Thm: CoDAG is unique minimizer of F}, the CoDAG equilibrium $\bar \arcLoadMod^\beta(\toll)$ is the unique minimizer of the strictly convex function $F(\cdot, \toll): \arcsLoadConstraintSet \ra \R$ defined by \eqref{Eqn: Def, F}. Thus, $\bar \arcLoadMod^\beta(\toll)$ can be characterized by the first-order optimality conditions of this optimization problem. This in turn allows us to establish
monotonicity.
For details, please see Appendix \ref{subsubsec: A1, Proof of Lemma: Monotonicity of CoDAG Equilibrium} \cite{Chiu2023DynamicTollingInArcBasedTAMs}.
\end{proof}

We then use the above lemmas to prove that the fixed-point equation (\ref{Eqn: Toll, Fixed Point Equation}) yields a unique solution.

\begin{lemma} \label{Lemma: p bar uniquely exists}
There exists a unique $\bar \toll \in \R^{|A_O|}$ satisfying \eqref{Eqn: Toll, Fixed Point Equation}:
\begin{align*}
    \bar \toll_{[a]} = \bar w^{\beta}_{[a]}(\bar \toll) \cdot \frac{d\latency_{[a]}}{d w} \big(\bar w^{\beta}_{[a]}(\bar \toll) \big), \hspace{1cm} \forall \hspace{0.5mm} [a] \in A_O. 
\end{align*}
\end{lemma}

\vspace{2mm}
\begin{proof}(\textbf{Proof Sketch})
Existence follows from the Brouwer fixed point theorem, since $\bar \arcLoadMod^\beta(\toll)$ is continuous in $\toll$ (Lemma \ref{Lemma: C1 of CoDAG Equilibrium}). Uniqueness follows via a contradiction argument; we show that the existence of two distinct fixed points of \eqref{Eqn: Toll, Fixed Point Equation} would violate the monotonicity established by Lemma \ref{Lemma: Monotonicity of CoDAG Equilibrium}.
For details, please see Appendix \ref{subsubsec: A1, Proof of Lemma: p bar uniquely exists} \cite{Chiu2023DynamicTollingInArcBasedTAMs}.
\end{proof}

Finally, we prove that the CoDAG equilibrium flow corresponding to $\bar \toll \in \R^{|\arcsOrig|}$ is perturbed socially optimal.

\begin{lemma}
\label{Lemma: p bar is Perturbed Socially Optimal}
$\bar w^\beta(\bar \toll)$ is perturbed socially optimal.
\end{lemma}

\vspace{1mm}
\begin{proof}(\textbf{Proof Sketch})
This follows by comparing the KKT conditions satisfied by $\bar \arcLoadMod^\beta(\bar \toll)$ (Theorem \ref{Thm: CoDAG is unique minimizer of F}) with the KKT conditions of the optimization problem that defines the perturbed socially optimal flow in Definition \ref{Def: Perturbed Socially Optimal Flow}.
For details, please see Appendix \ref{subsubsec: A1, Proof of Lemma: p bar is Perturbed Socially Optimal} \cite{Chiu2023DynamicTollingInArcBasedTAMs}.
\end{proof}

Together, Lemmas \ref{Lemma: C1 of CoDAG Equilibrium}, \ref{Lemma: Monotonicity of CoDAG Equilibrium}, \ref{Lemma: p bar uniquely exists}, and \ref{Lemma: p bar is Perturbed Socially Optimal} prove Theorem \ref{Thm: Fixed Point, p}.

\section{DYNAMICS AND CONVERGENCE}
\label{sec: Dynamics, Convergence}

\subsection{Discrete-time Dynamics}
\label{subsec: Discrete-time Dynamics}
Here, we present discrete-time stochastic dynamics that describes the evolution of the traffic flow and tolls on the network. Formally, $g_o$ 
units of traveler flow enter the network at the origin node $o$ at each time step $n \geq 0$. At each non-destination node $i \in \nodesMod \backslash \{d\}$, a \(\probDist_a[n]\) fraction of travelers chooses an outgoing arc  $a \in \arcsMod_i^+$. We shall refer to \(\probDist_a[n]\) as the \textit{aggregate arc selection probability}. Consequently, the flow induced on any arc \(a\in A\) satisfies:
\begin{align} \label{Eqn: General Network, w flow, discrete}
\arcLoadModDiscrete_a[n] = \Bigg( g_{i_a} + \sum_{a' \in \arcsMod_{i_a}^+} \arcLoadModDiscrete_{a'}[n] \Bigg) \cdot \probDist_a[n].
\end{align}
% ~\\ \frank{To edit below} ~\\
At the conclusion of every time step \(n\), travelers reach the destination node \(d\) and observe a noisy estimate of the cost-to-go values and tolls on all arcs in the network (including arcs not traversed during that time step).
Let $K_i > 0$ denote node-dependent constants, and let $\{\eta_i[n+1] \in \R: i \in \nodesMod, n \geq 0 \}$ be independent bounded random variables\footnote{The random variables $\{\eta_a[n]: a \in \arcsMod, n \geq 0\}$ are assumed to be independent of travelers' perception uncertainties.} in $[\underline \mu, \overline \mu]$, with $0 < \underline \mu < \mu < \overline \mu < 1/\max\{K_i: i \in \nodesMod \backslash \{d\}\}$ and $\E[\eta_{i_a}[n+1]] = \mu$ at each node $i \in I$ and discrete time index $n \geq 0$. 
At the next time $n+1$ and non-destination node $i\in \nodesMod \backslash\{d\}$, a $\eta_{i}[n+1] \cdot K_i$ fraction of travelers at node $i \in \nodesMod$ observes the latencies on each arc, and decides to switch to the outgoing arc that minimizes the (stochastic) observed cost-to-go. Meanwhile, $1-\eta_{i}[n+1] \cdot K_i$ fraction of travelers selects the same arc they used at time step $n$. 
% \young{What does the next sentence want to convey?} Observe that the cost-to-go on each arc $[a] \in \arcsOrig$ is an increasing function of the congestion on that arc $\arcLoadModDiscrete_{[a]}[n] := \sum_{a' \in [a]} \arcLoadModDiscrete_{a'}[n]$, which in turn depend on travelers' aggregate arc selection decisions (see \eqref{Eqn: General Network, w flow, discrete}). 
% At the end of every time step $n$, every traveler observe a noisy estimate of the latency to go, $\tilde{z}_a$, from every arc $a$ on $G$.
Thus, the arc selection probabilities evolve according to the following \textit{perturbed best-response dynamics}:
\begin{align} \label{Eqn: General Network, xi flow, discrete}
    &\probDist_a[n+1] 
    % \tag{\textsf{\(\probDist\)-update}} 
    \\ \nonumber
    = \hspace{0.5mm}  &\probDist_a[n] + \eta_{i_a}[n+1] \cdot K_{i_a} \\ \nonumber
    &\hspace{2mm} \cdot \Bigg(-\probDist_a[n] + \frac{\exp(-\beta \big[ \costToGo_a(\arcLoadModDiscrete[n], \tollDiscrete[n]) \big])}{\sum_{ a' \in \arcsMod_{i_a}^+} \exp(-\beta \big[ \costToGo_{a'}(\arcLoadModDiscrete[n], \tollDiscrete[n]) \big])} \Bigg). 
\end{align}
We assume that $\probDist_a[0] > 0$ for each $a \in \arcsMod$, i.e., each arc has some strictly positive initial traffic flow. This captures the stochasticity in travelers' perception of network congestion that causes each arc to be assigned a nonzero probability of being selected. 

At each time step $n+1 \geq 0$, the tolls $\tollDiscrete_{[a]}[n] \in \R^{|\arcsOrig|}$ on each arc $[a] \in \arcsOrig$ are updated by interpolating between the tolls implemented at time step \(n\), and the marginal latency of that arc given the flow at time step \(n\). That is:
\begin{align}
    \label{Eqn: General Network, toll flow, discrete} 
    &\tollDiscrete_{[a]}[n+1] \\ \nonumber
    = \hspace{0.5mm} &\tollDiscrete_{[a]}[n] + \gamma\left( - \tollDiscrete_{[a]}[n] + \arcLoadModDiscrete_{[a]}[n] \cdot \frac{d\latency_{[a]}}{d\arcLoadMod}(\arcLoadModDiscrete_{[a]}[n]) \right),
\end{align}
with $\gamma \in (0, 1)$\footnote{Our result also holds if $\gamma$ is a random variable with bounded support.}, where with a slight abuse of notation, we denote $\arcLoadModDiscrete_{[a]} := \sum_{a' \in [a]} \arcLoadModDiscrete_{a'}$. 
% We note that our results in this work also hold if \(\gamma\) is a random variable with bounded support.  
% In words, the updated toll on each arc linearly interpolates the current toll and the marginal latency of that arc given the current traffic flow. 
Note that the update \eqref{Eqn: General Network, toll flow, discrete} is distributed, i.e., for each arc in the original network, the updated toll depends only on the flow of that arc, and not on the flow of any other arc. Moreover, we assume that $\gamma \ll \mu$, i.e., the toll updates \eqref{Eqn: General Network, toll flow, discrete} occur at a slower timescale compared to the arc selection probability updates \eqref{Eqn: General Network, xi flow, discrete}.

To simplify our study of the convergence of the dynamics \eqref{Eqn: General Network, xi flow, discrete} and \eqref{Eqn: General Network, toll flow, discrete}, we assume that the arc latency functions are affine in the congestion on the link.

\begin{assumption}\label{assm: AffineLatency}
Each arc latency function $\latency_{[a]}$ is affine, i.e.,: 
\begin{equation}
\label{eq: affine latency}
\latency_{[a]}(\arcLoadMod_{[a]}) = \theta_{\tilde a, 1} \arcLoadMod_{[a]} + \theta_{[a], 0},
\end{equation}
for some $\theta_{[a], 1}, \theta_{[a], 0} > 0$. 
\end{assumption}

Under Assumption \ref{assm: AffineLatency}, the toll dynamics \eqref{Eqn: General Network, toll flow, discrete} can be alternatively written as follows 
\begin{align}\label{Eqn: Actual Toll Update}
    P_{[a]}[n+1]=\tollDiscrete_{[a]}[n] + \gamma\left( - \tollDiscrete_{[a]}[n] + \arcLoadModDiscrete_{[a]}[n] \cdot \theta_{[a], 1} \right). 
    % \tag{\textsf{\(P\)-update}}
\end{align}
\subsection{Convergence Results}
\label{subsec: Convergence Results}
% Recall that Theorem \ref{Thm: Fixed Point, p} establishes the existence and uniqueness of an \textit{optimal toll} $\bar \toll \in \R^{|\arcsOrig|}$ satisfying \eqref{Eqn: Toll, Fixed Point Equation}. 
In this subsection, we show that the arc selection probability and toll updates \eqref{Eqn: General Network, xi flow, discrete}-\eqref{Eqn: Actual Toll Update} converge in the neighborhood of the socially optimal flow $\bar \arcLoadMod^\beta(\bar \toll)$ and the corresponding toll $\bar \toll$ respectively.
% he optimal toll values and corresponding CoDAG equilibrium,
 % (Theorem \ref{Thm: Convergence, w, p, discrete}). 
% Moreover, if the latency function $\latency_{[a]}: [0, \infty) \ra [0, \infty)$ is affine for each arc $[a] \in \arcsOrig$, the toll and flow dynamics converge globally (Corollary \ref{Cor: Global Convergence, w, p, discrete}).
% \begin{theorem}
    
% \end{theorem}
% \begin{theorem} \label{Thm: Convergence, w, p, discrete}
% There exists some $\epsilon > 0$ such that, if $\Vert \tollDiscrete[0] - \bar \toll \Vert_2 \leq \epsilon$, then (a):
% {\small
% \begin{align*}
%     \limsup_{n \ra \infty} \E\big[ \Vert \probDist[n] - \bar \probDist^\beta(\bar \toll) \Vert_2^2 + \Vert \tollDiscrete[n] - \bar \toll \Vert_2^2 \big] = O\left(\mu + \frac{a}{\mu} \right),
% \end{align*}
% }
% % \begin{align*}
% %     &\limsup_{n \ra \infty} \E\big[ \Vert \probDist[n] - \bar \probDist^\beta(\bar \toll) \Vert_2^2 + \Vert \tollDiscrete[n] - \bar \toll \Vert_2^2 \big] \\
% %     = \hspace{0.5mm} &O\left(\mu + \frac{a}{\mu} \right),
% % \end{align*}
% and (b) for each $\delta > 0$:
% {\small
% \begin{align*}
%     &\limsup_{n \ra \infty} \Prob \big[ \Vert \probDist[n] - \bar \probDist^\beta(\bar \toll) \Vert_2^2 + \Vert \tollDiscrete[n] - \bar \toll \Vert_2^2 \geq \delta \big]\\
%     = \hspace{0.5mm} &O\left(\frac{\mu}{\delta} + \frac{a}{\delta \mu} \right).
% \end{align*}
% }
% \end{theorem}

\vspace{2mm}
\begin{theorem}\label{Thm: Convergence, w, p, discrete}
% Suppose $s_{[a]}: \arcsLoadConstraintSet \ra \R$ is affine for each $a \in \arcsMod$. Then (a):.
The joint evolution of arc selection probability and toll updates \eqref{Eqn: General Network, xi flow, discrete}-\eqref{Eqn: Actual Toll Update} satisfies
% \begin{align*}
%     &\limsup_{n \ra \infty} \E\big[ \Vert \probDist[n] - \bar \probDist^\beta(\bar \toll) \Vert_2^2 + \Vert \tollDiscrete[n] - \bar \toll \Vert_2^2 \big] \\
%     = \hspace{0.5mm} &O\left(\mu + \frac{a}{\mu} \right),
% \end{align*}
{
% \small
\normalsize
\begin{align*}
    &\limsup_{n \ra \infty} \E\big[ \Vert \probDist[n] - \bar \probDist^\beta(\bar \toll) \Vert_2^2 + \Vert \tollDiscrete[n] - \bar \toll \Vert_2^2 \big] \\
    = \hspace{0.5mm} &O\left(\mu + \frac{\gamma}{\mu} \right).
\end{align*}
}
Consequently, for each $\delta > 0$:
{
% \small
\normalsize
\begin{align*}
    &\limsup_{n \ra \infty} \Prob \big( \Vert \probDist[n] - \bar \probDist^\beta(\bar \toll) \Vert_2^2 + \Vert \tollDiscrete[n] - \bar \toll \Vert_2^2 \geq \delta \big) \\
    = \hspace{0.5mm} &O\left(\frac{\mu}{\delta} + \frac{\gamma}{\delta \mu} \right).
\end{align*}
}
\end{theorem}

\vspace{2mm}
To prove Theorem \ref{Thm: Convergence, w, p, discrete}, we employ the theory of two-timescale stochastic approximation \cite{Borkar2008StochasticApproximation}. Consequently, the asymptotic behavior of \eqref{Eqn: General Network, xi flow, discrete}-\eqref{Eqn: Actual Toll Update} can be characterized by studying the convergence properties of the corresponding continuous-time dynamical system. 
% This entails showing
% that the following continuous-time counterpart to the
% discrete-time updates \eqref{Eqn: General Network, xi flow, discrete} converges to the CoDAG equilibrium. 
Since the tolls are updated at a slower rate compared to the traffic flows ($\gamma \ll \mu$), we consider the evolution of continuous-time flows $\arcLoadMod(t)$ under a fixed toll $\toll \in \R^{|\arcsOrig|}$, and continuous-time tolls $p(t)$ with flow converged at the corresponding CoDAG equilibrium $\bar \arcLoadMod^\beta(\toll(t))$ at each time. Specifically, for any fixed toll $\toll \in \R^{|\arcsOrig|}$, on each arc $a \in \arcsMod$, the arc selection probabilities evolve as follows:
{\small
\begin{align} \label{Eqn: General Network, w flow, continuous}
    \arcLoadMod_a(t) &= \probDist_a(t) \cdot \Bigg(\nodeLoadIn_{i_a} + \sum_{a' \in \arcsMod_{i_a}^-} \arcLoadMod_{a'}(t)\Bigg), \\ \label{Eqn: General Network, xi flow, continuous}
    \dot \probDist_a(t) &= K_{i_a} \cdot \left( - \probDist_a(t) + \frac{\exp(-\beta \cdot \costToGo_a(\arcLoadMod(t), \toll))}{\sum_{ a' \in \arcsMod_{i_a}^+} \exp(-\beta \cdot \costToGo_{a'}(\arcLoadMod(t), \toll))} \right).
\end{align}
}
Meanwhile, on each arc $[a] \in \arcsOrig$ in the original network, we consider the following continuous-time toll dynamics:
\begin{align} \label{Eqn: General Network, p flow, continuous}
    \dot \toll_{[a]}(t) &= - \toll_{[a]}(t) + \bar \arcLoadMod_{[a]}^\beta(\toll(t)) \cdot \theta_{[a],1}.
\end{align}

We prove that, for each fixed toll $\toll \in \R^{|\arcsOrig|}$, the corresponding continuous-time $\probDist$-dynamics \eqref{Eqn: General Network, xi flow, continuous} globally asymptotically converges to the corresponding CoDAG equilibrium $\bar \arcLoadMod^\beta(\toll) \in \R^{|\arcsMod|}$. Moreover, the continuous-time toll dynamics \eqref{Eqn: General Network, p flow, continuous} globally converges to the optimal toll $\bar \toll \in \R^{|\arcsOrig|}$.

\begin{lemma}[\textbf{Informal}] \label{Lemma: Convergence, w, continuous}
Suppose $\arcLoadMod(0) \in \arcsLoadConstraintSet$, i.e., the initial flow satisfies flow continuity. Under the continuous-time flow dynamics \eqref{Eqn: General Network, xi flow, continuous} and \eqref{Eqn: General Network, w flow, continuous}, if $K_i \ll K_{i'}$ whenever $\ell_i < \ell_{i'}$, the continuous-time traffic allocation $\arcLoadMod(t)$ globally asymptotically converges to the corresponding CoDAG equilibrium $\bar \arcLoadMod^\beta(\toll)$.
\end{lemma}

\begin{proof}(\textbf{Proof Sketch}) 
The following proof sketch parallels that of \cite{Chiu2023ArcbasedTrafficAssignment}, Lemma 2, and is included for completeness.
Recall that Theorem \ref{Thm: CoDAG is unique minimizer of F} establishes $\bar \arcLoadMod^\beta(\toll)$ as the unique minimizer of the map $F(\cdot, \toll): \arcsLoadConstraintSet \ra \R$, defined by \eqref{Eqn: Def, F}. We show that $F(\cdot, \toll)$ is a Lyapunov function for the continuous-time flow dynamics induced by \eqref{Eqn: General Network, xi flow, continuous}. To this end, we first unroll the dynamics \eqref{Eqn: General Network, xi flow, continuous} using \eqref{Eqn: General Network, w flow, continuous}, as follows:
{\small
\begin{align*}
    &\hspace{5mm} \dot\arcLoadMod_a(t) \\
    &= - K_{i_a} \cdot \Bigg( 1 - \frac{1}{K_{i_a}} \cdot \frac{ \sum_{a' \in \arcsMod_{i_a}^-} \dot \arcLoadMod_{a'}(t)}{\sum_{\hat a \in \arcsMod_{i_a}^+} \arcLoadMod_{\hat a}(t)} \Bigg) \arcLoadMod_a(t) \\ \nonumber
    &\hspace{5mm} + K_{i_a} \cdot \sum_{a' \in \arcsMod_{i_a}^-} \arcLoadMod_{a'}(t) \cdot \frac{\exp(-\beta \costToGo_a(\arcLoadMod(t), \toll))}{\sum_{a' \in \arcsMod_{i_a}^+} \exp(-\beta \costToGo_{a'}(\arcLoadMod(t), \toll))}.
\end{align*}
}
Next, we establish that if $\arcLoadMod(0) \in \arcsLoadConstraintSet$, then for each $t \geq 0$:
\begin{align*}
    \dot F(t) &= \dot w(t)^\top \nabla_w F(w(t)) \leq 0.
\end{align*}
The proof then follows from LaSalle's Theorem (see \cite[Proposition 5.22]{Sastry1999NonlinearSystems}). 
For a precise statement of Lemma \ref{Lemma: Convergence, w, continuous}, please see Appendix \ref{subsubsec: A2, Statement of Lemma: Convergence, w, continuous} \cite{Chiu2023DynamicTollingInArcBasedTAMs}; for the proof of the analogous theorem in \cite{Chiu2023ArcbasedTrafficAssignment}, please see \cite{Chiu2023ArcbasedTrafficAssignment} Appendix C.1.
\end{proof}

\begin{lemma} \label{Lemma: Convergence, p, continuous}
The continuous-time toll dynamics \eqref{Eqn: General Network, p flow, continuous} globally exponentially converges to the CoDAG equilibrium $\bar \arcLoadMod^\beta(\bar \toll)$ corresponding to the optimal toll $\bar \toll$.
% Moreover, if the latency function $\latency_{[a]}: \R^{|\arcsMod|} \ra \R$ is affine for each $a \in \arcsMod$, then the convergence is global.
\end{lemma}

\begin{proof} 
Define $D \in \R^{|\arcsOrig| \times |\arcsOrig|}$ to be the diagonal and symmetric positive definite matrix whose $[a]$-th diagonal element is given by:
\begin{align*}
    \frac{d \latency_{[a]}}{d\arcLoadMod}\big( \bar \arcLoadMod_{[a]}^\beta (\bar \toll) \big) = \theta_{[a], 1} > 0,
\end{align*}
for each $[a] \in \arcsOrig$.  Note that $D$ is independent of the toll $\toll$. Now, consider the Lyapunov function $V: \R^{|\arcsOrig|} \ra \R$, defined by:
\begin{align*}
    V(\toll) &:= \frac{1}{2} (\toll - \bar \toll)^\top D^{-1}(\toll - \bar \toll).
\end{align*}
The trajectory of the continuous-time toll dynamics \eqref{Eqn: General Network, p flow, continuous}, starting at $\toll(0)$, satisfies:
\begin{align*}
    &\hspace{5mm} \dot V(\toll(t)) \\
    &= (p(t)-\bar{p})^{\top}D^{-1}\dot{p}(t) \\ 
    &=\sum_{[a]\in A_O} \frac{(p_{[a]}(t)-\bar{p}_{[a]})}{\theta_{[a],1}} \cdot \left(-p_{[a]}(t)+\theta_{[a],1}\bar{w}^{\beta}_{[a]}(p(t))\right) \\
    &=\sum_{[a]\in A_O} \frac{(p_{[a]}(t)-\bar{p}_{[a]})}{\theta_{[a],1}}
    \\ 
    &\hspace{1cm} \cdot \left(-p_{[a]}(t)+\bar{p}_{[a]} -\bar{p}_{[a]}+\theta_{[a],1}\bar{w}^{\beta}_{[a]}(p(t))\right) \\
    &= - 2V(\toll(t)) \\
    &\hspace{1cm} +\sum_{[a]\in A_O}(p_{[a]}(t)-\bar{p}_{[a]})\left(\bar{w}_{[a]}^{\beta}(p(t)) - \bar{w}_{[a]}^{\beta}(\bar{p})  \right) \\
    &\leq -2V(\toll(t)),
\end{align*}
% This requires showing that $\frac{d \arcLoadMod^\beta}{d\toll}(\toll) \in \R^{|\arcsOrig| \times |\arcsOrig|}$ is continuous in, and symmetric negative semidefinite at, each $\toll \in \R^{|\arcsOrig|}$ (Lemma \ref{Lemma: C1 of CoDAG Equilibrium}).
where the final inequality follows due to the monotonicity of the map $\bar \arcLoadMod^\beta(\cdot)$ (Lemma \ref{Lemma: Monotonicity of CoDAG Equilibrium}). 
\end{proof}

To conclude the proof of Theorem \ref{Thm: Convergence, w, p, discrete}, it remains to check that the discrete-time dynamics \eqref{Eqn: General Network, xi flow, discrete}-\eqref{Eqn: Actual Toll Update}, and the continuous-time dynamics \eqref{Eqn: General Network, xi flow, continuous}-\eqref{Eqn: General Network, p flow, continuous}, satisfy the technical conditions in Lemmas \ref{Lemma: Technical Conditions for Stochastic Approximation, 1} and \ref{Lemma: Technical Conditions for Stochastic Approximation, 2}. In particular, Lemma \ref{Lemma: Technical Conditions for Stochastic Approximation, 1} establishes that flows and tolls are uniformly bounded across the arc and time indices, while Lemma \ref{Lemma: Technical Conditions for Stochastic Approximation, 2} asserts that the continuous-time flow and toll dynamics maps are Lipschitz continuous. 
% Finally, Corollary \ref{Cor: Global Convergence, w, p, discrete} follows directly by specializing the arguments in the proof of Theorem \ref{Thm: Convergence, w, p, discrete}. In particular, we invoke the global convergence of the continuous-time toll dynamics under the affine latency assumption, as established by Lemma \ref{Lemma: Convergence, p, continuous}.

\begin{lemma} \label{Lemma: Technical Conditions for Stochastic Approximation, 1}
The continuous-time flow and toll dynamics induced by \eqref{Eqn: General Network, xi flow, discrete}-\eqref{Eqn: Actual Toll Update} satisfy:
\begin{enumerate}
    \item For each $a \in \arcsMod$: $\{M_a[n+1]: n \geq 0\}$ is a martingale difference sequence with respect to the filtration $\mathcal{F}_n := \sigma\big( \cup_{a \in \arcsMod} (\arcLoadModDiscrete_a[1], \probDist[1], \toll[1], \cdots, \arcLoadModDiscrete_a[n], \probDist[n], \toll[n]) \big)$.
    
    \item There exist $C_w, C_m, C_p > 0$, independent of the node-dependent values $\{K_i: i \in \nodesMod\}$, such that, for each $a \in \arcsMod$ and each $n \geq 0$, we have $\arcLoadModDiscrete_a[n] \in [C_w, g_o]$, $\tollDiscrete_a[n] \in [0, C_p]$, and $|M_a[n]| \leq C_m$.
    % \begin{align*}
    %     \arcLoadModDiscrete_a[n] &\in [C_w, g_o], \\
    %     \tollDiscrete_a[n] &\in [0, C_p], \\
    %     |M_a[n]| &\leq C_m.
    %     % |\costToGo_a(\arcLoadModDiscrete[n], \tollDiscrete[n])| &\leq C_z, \\
    %     % |h_a(\arcLoadModDiscrete_a[n], \tollDiscrete_a[n])| &\leq C_h, \\
    % \end{align*}
    % Thus, $\E[\Vert \arcLoadModDiscrete[n] \Vert_2^2] < \infty$, $\E[\Vert \tollDiscrete[n] \Vert_2^2] < \infty$, and $\E[\Vert \tollDiscrete[n] \Vert_2^2] < \infty$.
\end{enumerate}
Likewise, the continuous-time flow and toll dynamics induced by \eqref{Eqn: General Network, xi flow, continuous} and \eqref{Eqn: General Network, p flow, continuous} satisfy:
\begin{enumerate}[resume]
    \item For each $a \in \arcsMod$, $t \geq 0$, we have $\arcLoadMod_a(t) \in [C_w, g_o]$ and $\toll_a(t) \in [0, C_p]$.
    % :
    % \begin{align*}
    %     \arcLoadMod_a(t) &\in [C_w, g_o], \\
    %     \toll_a(t) &\in [0, C_p].
    % \end{align*}
\end{enumerate}
\end{lemma}

\begin{proof}
Please see Appendix \ref{subsubsec: A2, Proof of Lemma: Technical Conditions for Stochastic Approximation, 1} \cite{Chiu2023DynamicTollingInArcBasedTAMs}.
\end{proof}

\begin{lemma} \label{Lemma: Technical Conditions for Stochastic Approximation, 2}
The continuous-time flow dynamics \eqref{Eqn: General Network, w flow, continuous} and toll dynamics \eqref{Eqn: General Network, p flow, continuous} satisfy:
\begin{enumerate}
    % \item There exists some $C_w' > 0$ such that $\arcLoadMod_a(t) \leq C_w'$ for each $a \in \arcsMod$ and $t \geq 0$.

    \item The map $\bar \probDist^\beta: \R^{|\arcsOrig|} \ra \R^{|\arcsMod|}$ is Lipschitz continuous.
    
    \item For each $a \in \arcsMod$, the restriction of the cost-to-go map $\costToGo_a: \arcsLoadConstraintSet \times \R^{|\arcsOrig|} \ra \R$ to the set of realizable flows and tolls, i.e., $\arcsLoadConstraintSet' \times [0, C_p]^{|\arcsOrig|}$, is Lipschitz continuous.
    
    \item The map from the probability transitions $\probDist \in \prod_{i \in I \backslash \{d\}} \Delta(\arcsMod_i^+)$ and the traffic flows $\arcLoadMod \in \arcsLoadConstraintSet$ is Lipschitz continuous.
    
    \item For each $a \in \arcsMod$, the restriction of the continuous dynamics transition map $\rho_a: \R^{|\arcsMod|} \times \R^{|\arcsOrig|} \ra \R^{|\arcsMod|}$, defined recursively as follows for each $a \in \arcsMod$:
    \begin{align*}
        \rho_a(\probDist, \toll) &:= - \probDist_a
        + \frac{\exp(-\beta \costToGo_a(\arcLoadMod, \toll))  }{\sum_{ a' \in \arcsMod_{i_a}^+} \exp(-\beta \costToGo_{a'}(\arcLoadMod, \toll))}
    \end{align*}
    to the set of realizable flows and tolls, i.e., $\arcsLoadConstraintSet' \times [0, C_p]^{|\arcsOrig|}$, is Lipschitz continuous.
    
    \item For each $a \in \arcsMod$, the map $r_{[a]}: \R^{|\arcsOrig|} \times \R^{|\arcsOrig|}$, defined as follows for each $a \in \arcsMod$:
    \begin{align*}
        r_{[a]}(\toll) &:= - \toll_{[a]} + \bar \arcLoadMod_{[a]}^\beta(\toll) \cdot \frac{d \latency_{[a]}}{d\arcLoadMod}(\bar \arcLoadMod_{[a]}^\beta(\toll)),
    \end{align*}
    is Lipschitz continuous.
\end{enumerate}
\end{lemma}

\begin{proof}
Please see Appendix \ref{subsubsec: A2, Proof of Lemma: Technical Conditions for Stochastic Approximation, 2} \cite{Chiu2023DynamicTollingInArcBasedTAMs}.
\end{proof}

\section{EXPERIMENT RESULTS}
\label{sec: Results}

This section presents experiments that validate the theoretical convergence results of Section \ref{sec: Dynamics, Convergence}. We present simulation results illustrating that, under \eqref{Eqn: General Network, xi flow, discrete}-\eqref{Eqn: Actual Toll Update}, the traffic flows and tolls converge to a neighborhood of the socially optimal values, as claimed by Theorem \ref{Thm: Convergence, w, p, discrete}.

%In this section, we conduct numerical experiments to validate the theoretical analysis presented in Section \refsec: Dynamics, Convergence. 
%We show in simulation that, under \eqref{Eqn: General Network, xi flow, discrete}, the traffic flows converge to a neighborhood of the condensed DAG equilibrium, as claimed by Theorem \ref{Thm: Convergence, w, discrete}.

\begin{figure}
    \centering
    \includegraphics[scale=0.7]{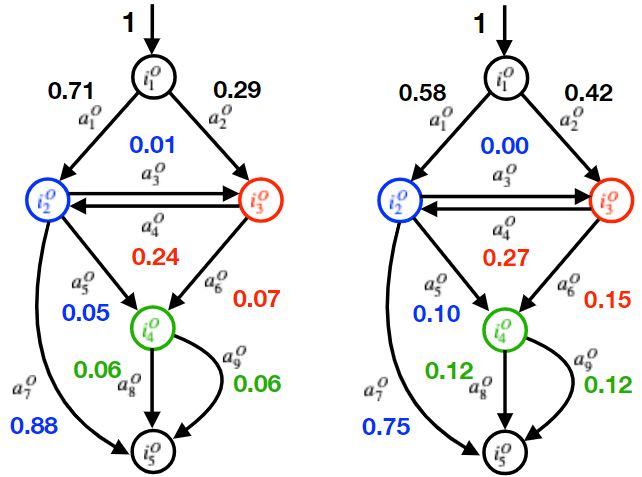}
    \caption{Steady state traffic flow on each arc for the original network before (left) and after (right) tolls. Flows on arcs emerging from the same node are represented in the same color.}
\label{fig:finalVer2}
\end{figure}

\begin{figure}
    \centering
    \includegraphics[scale=0.35]{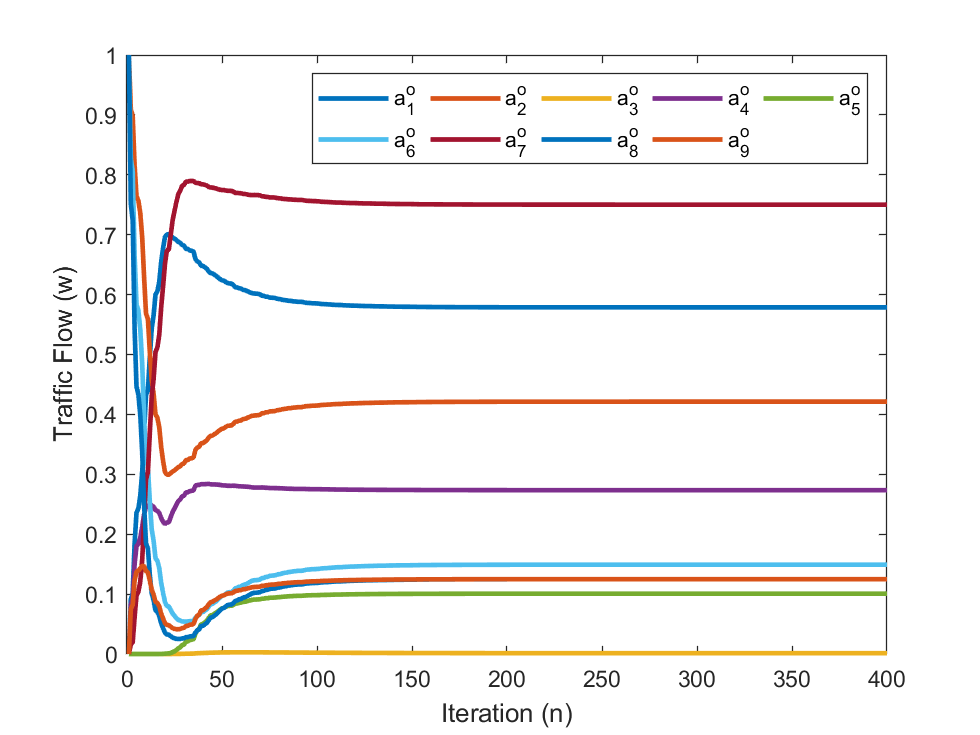}
    \caption{Traffic flow $W$ vs. time index $n$ for the condensed DAG in Figure \ref{fig:Front_Figure___Equivalent_DAG}.}
    \label{fig:w_dynamics}
\end{figure}

\begin{figure}
    \centering
    \includegraphics[scale=0.35]{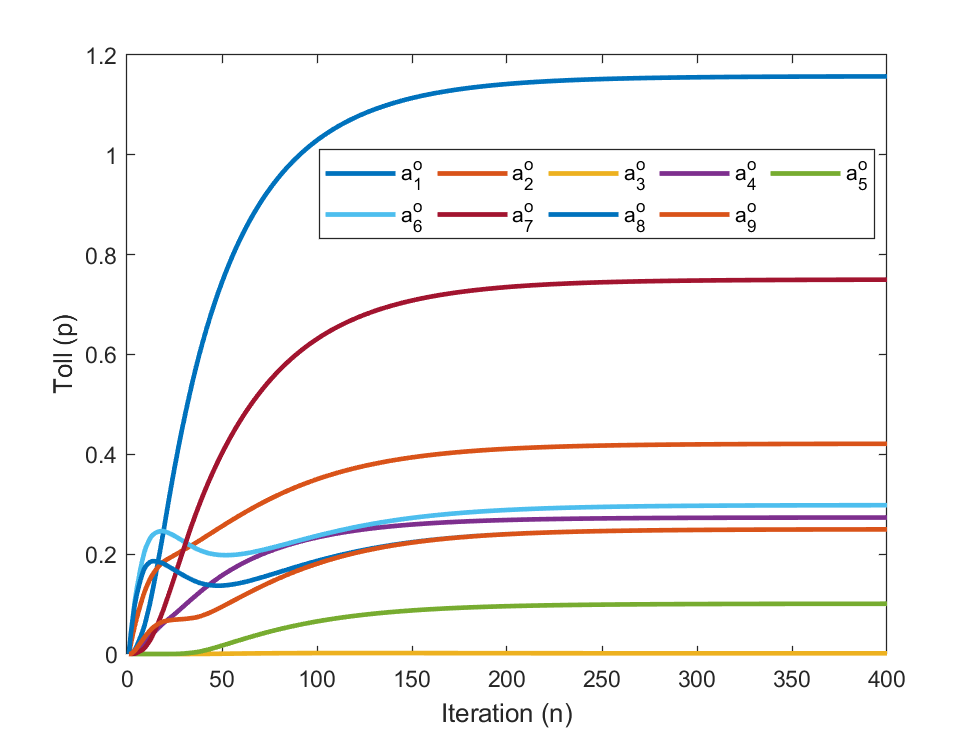}
    \caption{Toll $P$ vs. time index $n$ for the condensed DAG in Figure \ref{fig:Front_Figure___Equivalent_DAG}.}
    \label{fig:p_dynamics}
\end{figure}

Consider the network presented in Figure \ref{fig:Front_Figure___Equivalent_DAG}, following affine latency functions (\ref{eq: affine latency}) with parameters given in Table \ref{table: Parameters for simulation}. To validate Theorem \ref{Thm: Convergence, w, p, discrete}, we evaluate and plot the traffic flow values $W_a[n]$ and toll values $P_a[n]$ on each arc $a \in \arcsMod$ with respect to discrete time index $n \geq 0$. Figure \ref{fig:finalVer2} presents traffic flow values at the condensed DAG equilibrium (i.e., $\arcLoadMod^\beta$) for the original network before and after tolls. Meanwhile, Figure \ref{fig:w_dynamics} and \ref{fig:p_dynamics} illustrate that $w$ and $p$ converge to the condensed DAG equilibrium in approximately 300 iterations. As in \cite{Chiu2023ArcbasedTrafficAssignment}, flow convergence to the optimal allocation occurs even when the constants $\{K_i: i \in \nodesMod \}$ are simply all set to 1. While the original traffic distribution is more concentrated on a few routes, tolls can distribute the traffic more evenly. This shows that tolls can improve overall social welfare by reducing congestion in over-utilized routes. 

%While travelers generally prefer routes of lower latency, each route has a nonzero level of traffic flow at equilibrium. The reason is that under the perturbed best response dynamics, users do not allocate all the traffic flow to the minimum-cost route, but instead distribute their traffic allocation more evenly. 

\begin{table}
\centering\footnotesize
\caption{Parameters for simulation.}
% \vspace{-0.5cm}
\label{table: Parameters for simulation}\def\arraystretch{.9}
\begin{tabular}[t]{p{1cm}p{5.5cm}}
\toprule
{\bf Notation} & {\bf Default value}\\
\midrule
$\theta_{\tilde a, 0}$ & 0, 1, 0, 1, 1, 0, 1, 1, 1 (ordered by edge index)\\
$\theta_{\tilde a, 1}$ & 2, 1, 1, 1, 1, 1, 2, 2, 2 (ordered by edge index) \\
$g_1$ & 1\\
$\beta$ & 10\\
$\gamma$ & $0.02$\\
$\eta_{i_a}[n]$ & Uniform($0, 0.1$), $\forall a \in \arcsMod, i \in \nodesMod \backslash \{d\}$\\
\bottomrule
\end{tabular}
\end{table}

% \begin{figure}
%     \centering
%     \includegraphics[scale=0.5]{Figures/wpz.png}
%     \caption{}
%     \label{fig:wpz}
% \end{figure}

% \begin{figure}
%     \centering
%     \includegraphics[scale=0.22]{Figures/z_dynamics.png}
%     \caption{Cost-to-go $z$ vs. time index $n$ for the condensed DAG in Figure \ref{fig:Front_Figure___Equivalent_DAG}.}
%     \label{fig:z_dynamics}
% \end{figure}

%~\\
%\frank{INCOMPLETE; to be filled in.}
%~\\

\section{CONCLUSION AND FUTURE WORK}
\label{sec: Conclusion and Future Work}

This work introduces a discrete-time adaptive tolling scheme to minimize the total travel latency in a general traffic network with bidirectional edges. Our model assumes that, at each time, players near-instantaneously react via perturbed best response to the announced tolls. Accordingly, we formulate a two-timescale stochastic dynamical system that describes the joint evolution of traffic flow and tolls. We prove that the fixed point of these dynamics is unique and corresponds to the optimal traffic flow allocation from the perspective of minimizing the total travel time. Moreover, we prove that the stochastic dynamics converges to a neighborhood of the unique fixed point with high probability. Finally, we present simulation results that corroborate our theoretical findings.

Interesting avenues of future research include: (1) Extending our theoretical analysis to the setting where the latency function of each arc is not necessarily affine, (2) Developing tolling dynamics for the setting in which the central authority must learn the network latency functions and entropy regularization parameter $\beta > 0$ while simultaneously implementing an adaptive tolling scheme that converges to the optimal toll, and (3) Designing robust tolls for traffic networks in which some fraction of the population behaves unexpectedly or adversarially.

\printbibliography

\appendix

Please use the following link to access the ArXiv version with the appendix \cite{Chiu2023DynamicTollingInArcBasedTAMs} (\url{https://arxiv.org/pdf/2307.05466.pdf}). The authors will make certain that this link stays active.
% An ArXiv version of this paper is also available.

% \input{A0_Brief_Depth_Height}

\newpage

Below, we present proofs omitted in the main paper due to space limitations.

\subsection{Proofs for Section \ref{sec: Optimal Toll: Existence and Uniqueness}}
\label{subsec: A1, Optimal Toll: Existence and Uniqueness}

Here, we provide the proofs of Lemmas \ref{Lemma: C1 of CoDAG Equilibrium}, \ref{Lemma: Monotonicity of CoDAG Equilibrium}, \ref{Lemma: p bar uniquely exists}, and \ref{Lemma: p bar is Perturbed Socially Optimal}.

\subsubsection{Proof of Lemma \ref{Lemma: C1 of CoDAG Equilibrium}}
\label{subsubsec: A1, Proof of Lemma: C1 of CoDAG Equilibrium}

% First, we present the proof of Lemma \ref{Lemma: C1 of CoDAG Equilibrium}, restated as follows: $\bar \arcLoadMod^\beta(\toll)$ is continuously differentiable in $\toll$, and is symmetric negative definite at each $\toll \in \R^{|\arcsOrig|}$.

% \begin{proof}
Define $F:\arcsLoadConstraintSet\times \R^{|A_O|}  \ra \R$ by:
{\small
\begin{align} \nonumber
    &F(\arcLoadMod, \toll) \\
    := \hspace{0.5mm} &\sum_{[a] \in A_O} \int_0^{\arcLoadMod_{[a]}} \big[ \latency_{[a]}(\costToGo) + \toll_{[a]} \big] \hspace{0.5mm} dz \\
    &+ \frac{1}{\beta} \sum_{i \ne d} \Bigg[ \sum_{a \in A_i^+} \arcLoadMod_a \ln \arcLoadMod_a - \Bigg(\sum_{a \in A_i^+} \arcLoadMod_a \Bigg) \ln \Bigg(\sum_{a \in A_i^+} \arcLoadMod_a \Bigg) \Bigg].
\end{align}
}
The theory of constrained optimization implies that, for each $\toll$, the unique minimizer of $F(\cdot,p): \arcsLoadConstraintSet \ra \R$ is completely characterized via a set of equality constraints, which we describe below. First, recall that since $\arcsLoadConstraintSet$ is a subset of an affine subspace of $\R^{|A|}$ characterized by $|\nodesMod \backslash \{d\}|$ equality constraints, there exist $M \in \R^{|A| \times |\nodesMod \backslash \{d\}|}$, of full column rank, and $b \in \R^{|\nodesMod \backslash \{d\}|}$ such that:
\begin{align*}
    \arcsLoadConstraintSet = \{w \in \R^{|A|}: M^\top w + b = 0, \arcLoadMod_a \geq 0, \hspace{0.5mm} \forall \hspace{0.5mm} a \in A \}.
\end{align*}
Moreover, by using QR decomposition, we can assume that the columns of $M$ are orthonormal. Next, let $B \in \R^{|A| \times (|A| - |\nodesMod \backslash \{d\}|)}$ be given such that the columns of $B$ have unit norm, are pair-wise orthogonal, and are each orthogonal to the subspace of $\R^{|A|}$ spanned by the columns of $M$, i.e., $B^\top$ maps each vector in $\R^{|A|}$ to the coefficients of its projection onto the linear subspace orthogonal to $\arcsLoadConstraintSet$, with respect to an ordered, orthonormal basis of that subspace. Then the theory of constrained optimization, and the strict convexity of $F(\cdot, p)$, imply that $\bar w^\beta(p)$, the unique minimizer of $F( \cdot, p)$, is completely characterized by the equations:
\begin{align*}
    &M^\top w + b = 0, \\
    &B^\top \nabla_w F(\arcLoadMod, \toll) = 0.
\end{align*}

To this end, define $J: \R^{|A_O|} \times \R^{|A|} \ra \R^{|A|}$ by:
\begin{align*}
    J(\arcLoadMod, \toll) &:= \begin{bmatrix}
        M^\top w + b \\
        B^\top \nabla_w F(\arcLoadMod, \toll)
    \end{bmatrix}.
\end{align*}

Note that $J$ is continuously differentiable almost everywhere, with:
\begin{align*}
    \frac{\partial J}{\partial w}(\arcLoadMod, \toll) &= \begin{bmatrix}
        M^\top \\
        B^\top \nabla_w^2 F(\arcLoadMod, \toll)
    \end{bmatrix} \in \R^{|A| \times |A|}.
\end{align*}
Suppose by contradiction that $\frac{\partial J}{\partial w}(\arcLoadMod, \toll) \in \R^{|A| \times |A|}$ is singular at some $(\arcLoadMod, \toll)$. Then $\frac{\partial J}{\partial w}(\arcLoadMod, \toll)^\top \in \R^{|A| \times |A|}$
% $= \begin{bmatrix}
   % M & \nabla_w^2 F(\arcLoadMod, \toll) B
%\end{bmatrix}$ 
lacks full column rank, i.e.:
\begin{align*}
    \text{dim}(R(M) + R(\nabla_w^2 F(\arcLoadMod, \toll) B) =&\text{rank}(\begin{bmatrix}
    M & \nabla_w^2 F(\arcLoadMod, \toll) B \end{bmatrix}) \\
    \leq \hspace{0.5mm} &|A|-1.
\end{align*}
By the Boolean formula for sums of vector spaces:
\begin{align*}
    &\text{dim}(R(M) \cap R(\nabla_w^2 F(\arcLoadMod, \toll) B) \\
    = \hspace{0.5mm} &\text{dim}(R(M)) + \text{dim}(R(\nabla_w^2 F(\arcLoadMod, \toll) B) \\
    &\hspace{5mm} - \text{dim}(R(M) + R(\nabla_w^2 F(\arcLoadMod, \toll) B) \\
    = \hspace{0.5mm} &\text{dim}(R(M)) + \text{dim}(R(B)) \\
    &\hspace{5mm} - \text{dim}(R(M) + R(\nabla_w^2 F(\arcLoadMod, \toll) B) \\
    \geq \hspace{0.5mm} &|A| - (|A| - 1) \\
    = \hspace{0.5mm} &1.
\end{align*}
Thus, there exists some nonzero vector $v \in R(M) \cap R(\nabla_w^2 F(\arcLoadMod, \toll) B)$. Since $v \in R(M)$, and the columns of $B$ are orthogonal to $R(M)$, we have $B^\top v = 0$. Meanwhile, since $v \in R(\nabla_w^2 F(\arcLoadMod, \toll) B)$, there exists some nonzero $\arcLoadMod \in \R^{|A|-d}$ such that $u = \nabla_w^2 F(\arcLoadMod, \toll) B \arcLoadMod$. Thus, we have:
\begin{align*}
    0 = B^\top u = B^\top \nabla_w^2 F(\arcLoadMod, \toll) B u,
\end{align*}
a contradiction, since the fact that $B^\top $ has full row rank and $\nabla_w^2 F(\arcLoadMod, \toll)$ is symmetric positive definite implies that $B^\top \nabla_w^2 F(\arcLoadMod, \toll) B$ is symmetric positive definite, and $u \ne 0$ by construction. This establishes that $\frac{\partial J}{\partial w}(\arcLoadMod, \toll) \in \R^{|A| \times |A|}$ is non-singular at each $(\arcLoadMod, \toll) \in \R^{|A_O|} \times \R^{|A|}$. The existence and continuity of $\frac{d\bar \arcLoadMod^\beta}{d\toll}(\toll)$ at each $\toll \in \R^{|\arcsOrig|}$ now follows from the Implicit Function Theorem.
% \end{proof}

\subsubsection{Proof of Lemma \ref{Lemma: Monotonicity of CoDAG Equilibrium}}
\label{subsubsec: A1, Proof of Lemma: Monotonicity of CoDAG Equilibrium}
In this subsection, we show that for any $\toll, \toll' \in \R^{|A_O|}$:
\begin{align*}
    \sum_{a \in A} \Big( \bar \arcLoadMod_a^\beta(p') - \bar \arcLoadMod_a^\beta(p) \Big) (\toll_{[a]}' - \toll_{[a]}) \leq 0.
\end{align*}

By Theorem \ref{Thm: CoDAG is unique minimizer of F}, $\bar w^\beta(p)$ is the unique minimizer, in $\arcsLoadConstraintSet$, of the following strictly convex function of $\arcLoadMod$:
{\small
\begin{align*}
    &\sum_{[a] \in A_O} \int_0^{\arcLoadMod_{[a]}} \big[ \latency_{[a]}(\costToGo) + \toll_{[a]} \big] \hspace{0.5mm} dz \\
    + \hspace{0.5mm} &\frac{1}{\beta} \sum_{i \ne d} \Bigg[ \sum_{a \in A_i^+} \arcLoadMod_a \ln \arcLoadMod_a - \Bigg(\sum_{a \in A_i^+} \arcLoadMod_a \Bigg) \ln \Bigg(\sum_{a \in A_i^+} \arcLoadMod_a \Bigg) \Bigg].
\end{align*}
}

Applying first-order conditions for optimality in constrained convex optimization, we obtain that, for each $\arcLoadMod^1 \in \arcsLoadConstraintSet$:
\begin{align*}
    &\sum_{a \in A} \Bigg[ \latency_{[a]}\big(\bar \arcLoadMod_{[a]}^\beta(p) \big) + \toll_{[a]} + \frac{1}{\beta} \ln\left( \frac{\bar \arcLoadMod_a^\beta (p)}{\sum_{a' \in A_{i_a}^+} \bar \arcLoadMod_{a'}^\beta (p)} \right) \Bigg] \\
    &\hspace{1cm} \cdot (\arcLoadMod_a^1 - \bar \arcLoadMod_a^\beta (p)) \geq 0.
\end{align*}
Similarly, for $\bar \arcLoadMod_{[a]}(p')$, we obtain that for each $\arcLoadMod^2 \in \arcsLoadConstraintSet$:
\begin{align*}
    &\sum_{a \in A} \Bigg[ \latency_{[a]}\big(\bar \arcLoadMod_{[a]}^\beta(p') \big) + \toll_{[a]}' + \frac{1}{\beta} \ln\left( \frac{\bar \arcLoadMod_a^\beta (p')}{\sum_{a' \in A_{i_a}^+} \bar \arcLoadMod_{a'}^\beta (p')} \right) \Bigg] \\
    &\hspace{1cm} \cdot (\arcLoadMod_a^2 - \bar \arcLoadMod_a^\beta (p') ) \geq 0.
\end{align*}
Taking $\arcLoadMod^1 := \bar \arcLoadMod_a^\beta (p')$, $\arcLoadMod^2 := \bar \arcLoadMod_a^\beta (p)$, and adding the above two inequalities, we have:
{\small
\begin{align*}
    0 &\leq \sum_{a \in A} \big( \bar \arcLoadMod_a^\beta(p') - \bar \arcLoadMod_a^\beta(p) \big) \\
    &\hspace{3mm} \cdot \Bigg[ \latency_{[a]}(\bar \arcLoadMod_{[a]}^\beta (p)) - \latency_{[a]}(\bar \arcLoadMod_{[a]}^\beta (p')) + \toll_{[a]} - \toll_{[a]}' \\ 
    &\hspace{5mm} + \frac{1}{\beta} \ln\Bigg( \frac{\bar \arcLoadMod_a^\beta(p)}{\sum_{a' \in A_{i_a}^+} \bar \arcLoadMod_{a'}^\beta(p)} \Bigg) - \frac{1}{\beta} \ln\Bigg( \frac{\bar \arcLoadMod_a^\beta(p')}{\sum_{a' \in A_{i_a}^+} \bar \arcLoadMod_{a'}^\beta(p')} \Bigg) \Bigg].
\end{align*}
}
Since the maps $\arcLoadMod_a \mapsto \latency_{[a]}(\arcLoadMod_{[a]})$ and $\arcLoadMod_a \mapsto \ln\big( \arcLoadMod_a / \sum_{a' \in A_{i_a}^+} \arcLoadMod_{a'} \big)$ are non-decreasing, by rearranging terms, we obtain:
\begin{align*}
    \sum_{a \in A} \Big( \bar \arcLoadMod_a^\beta(p') - \bar \arcLoadMod_a^\beta(p) \Big) (\toll_{[a]}' - \toll_{[a]}) \leq 0,
\end{align*}
as desired. Additionally, it also holds that 
\begin{align*}
    \sum_{[a] \in A_O} \Big( \bar \arcLoadMod_{[a]}^\beta(p') - \bar \arcLoadMod_{[a]}^\beta(p) \Big) (\toll_{[a]} - \toll_{[a]}) \leq 0.
\end{align*}

\subsubsection{Proof of Lemma \ref{Lemma: p bar uniquely exists}}
\label{subsubsec: A1, Proof of Lemma: p bar uniquely exists}
In this subsection, we show that there exists a unique $\bar \toll \in \R^{|A_O|}$ satisfying \eqref{Eqn: Toll, Fixed Point Equation}:
\begin{align*}
    \bar \toll_{[a]} = \bar w^{\beta}_{[a]}(\bar \toll) \cdot \frac{d\latency_{[a]}}{d w} \big(\bar w^{\beta}_{[a]}(\bar \toll) \big), \hspace{1cm} \forall \hspace{0.5mm} [a] \in A_O. 
\end{align*}

Define $\psi: \R^{|A_O|} \ra \R$ as:
\begin{align*}
    \psi_{[a]}(p) := \arcLoadMod_{[a]}(p) \cdot \frac{d\latency_{[a]}}{dw} \big(\arcLoadMod_{[a]}(p) \big), \hspace{1cm} \forall \hspace{0.5mm} [a] \in A_O.
\end{align*}
Since $\arcLoadMod_{[a]}(\cdot)$ is continuous (Lemma \ref{Lemma: C1 of CoDAG Equilibrium}), and $\latency_{[a]}$ is continuously differentiable, the map $\psi$ is continuous. Define the set:
\begin{align*}
    K := \left\{ y \in \R^{|A_O|}: y \succeq 0, \Vert y \Vert_1 \leq |\arcsOrig| \nodeLoadIn_o \max_{[a] \in A_O} \frac{d \latency_{[a]}}{dw}(\nodeLoadIn_o) \right\}.
\end{align*}
Observe that $K$ is a compact and convex subset of $\R^{|A_O|}$, and $\psi$ maps $K$ to $K$, since for any $p \in K$, we have $\psi_a(p) \geq 0$ for each $a \in \arcsMod$, and:
\begin{align*}
    \Vert \psi(p) \Vert_1 &= \sum_{a \in \arcsOrig} \psi_a(p) \\
    &= \sum_{a \in \arcsOrig} \bar \arcLoadMod_{[a]}(p) \cdot \frac{ds_{[a]}}{d\arcLoadMod}(\bar \arcLoadMod_{[a]}(p)) \\
    &\leq \max_{a \in \arcsOrig} \frac{ds_{[a]}}{d\arcLoadMod}(g_o) \cdot \sum_{a \in \arcsOrig} \bar \arcLoadMod_{[a]}(p) \\
    &\leq |\arcsOrig| g_o \cdot \max_{a \in \arcsOrig} \frac{ds_{[a]}}{d\arcLoadMod}(g_o).
\end{align*}

. Thus, by the Brouwer's fixed point theorem, there exists a fixed point $\bar \toll \in K \subset \R^{|A_O|}$ of $\psi$, i.e., there exists $\bar \toll \in \R^{|A_O|}$ satisfying \eqref{Eqn: Toll, Fixed Point Equation}, i.e.,:
\begin{align*}
    \bar \toll_{[a]} = \bar \arcLoadMod^\beta_{[a]}(\bar \toll) \frac{d\latency_{[a]}}{d w} (\bar \arcLoadMod^\beta_{[a]}(\bar \toll)),  \hspace{1cm} \forall \hspace{0.5mm} [a] \in A_O.
\end{align*}
Next, we show that $\bar \toll$ is unique up to Markovian Traffic Equilibrium on the original traffic network, i.e., any $\toll' \in \R^{|A_O|}$ satisfies \eqref{Eqn: Toll, Fixed Point Equation} if and only if $\bar \arcLoadMod_{[a]}^\beta(p') = \bar \arcLoadMod_{[a]}^\beta(\bar \toll)$ for each $a \in A$. To show this, suppose by contradiction that there exists some $\toll' \in \R^{|A_O|}$ satisfying \eqref{Eqn: Toll, Fixed Point Equation}, such that $\bar \arcLoadMod_{[a]}^\beta (p') \ne \bar \arcLoadMod_{[a]}^\beta (\bar \toll)$ for some $[a] \in A_O$. Then:
\begin{align*}
    &\bar \toll_{[a]} - \toll_{[a]}' \\
    = \hspace{0.5mm} &\bar \arcLoadMod_{[a]}^\beta(\bar \toll) \cdot \frac{d\latency_{[a]}}{dw} \big(\bar \arcLoadMod_{[a]}^\beta(\bar \toll) \big) - \bar \arcLoadMod_{[a]}^\beta(p') \cdot \frac{d\latency_{[a]}}{dw} \big(\bar \arcLoadMod_{[a]}^\beta(p') \big) \\
    = \hspace{0.5mm} &\Big[ \bar \arcLoadMod_{[a]}^\beta(\bar \toll) - \bar \arcLoadMod_{[a]}^\beta(p') \Big] \cdot \frac{d\latency_{[a]}}{dw} \big(\bar \arcLoadMod_{[a]}^\beta(\bar \toll) \big) \\
    &\hspace{1cm} + \bar \arcLoadMod_{[a]}^\beta(p') \cdot \Big[ \frac{d\latency_{[a]}}{dw} \big(\bar \arcLoadMod_{[a]}^\beta(\bar \toll) \big) - \frac{d\latency_{[a]}}{dw} \big(\bar \arcLoadMod_{[a]}^\beta(p') \big) \Big].
\end{align*}
Rearranging terms, and invoking the strict convexity and increasing nature of each $\latency_{[a]}$, and the fact that $\bar \arcLoadMod_{[a]}^\beta(\bar \toll) \ne \bar \arcLoadMod_{[a]}^\beta(p')$ for some $[a] \in A_O$, we obtain:
\begin{align*}
    &\sum_{a \in A} \Big[ \bar \arcLoadMod_a^\beta(\bar \toll) - \bar \arcLoadMod_a^\beta(p') \Big] (\bar \toll_{[a]} - \toll_{[a]}') \\
    = \hspace{0.5mm} &\sum_{[a] \in A_O} \Big[ \bar \arcLoadMod_{[a]}^\beta(\bar \toll) - \bar \arcLoadMod_{[a]}^\beta(p') \Big] (\bar \toll_{[a]} - \toll_{[a]}') \\
    = \hspace{0.5mm} &\sum_{[a] \in A_O} \Big[ \bar \arcLoadMod_{[a]}^\beta(\bar \toll) - \bar \arcLoadMod_{[a]}^\beta(p') \Big]^2 \cdot \frac{d\latency_{[a]}}{dw}\big( \bar \arcLoadMod_{[a]}^\beta(\bar \toll) \big) \\
    &\hspace{5mm} + \sum_{[a] \in A_O} \bar \arcLoadMod_{[a]}^\beta(p') \Big[ \bar \arcLoadMod_{[a]}^\beta(\bar \toll) - \bar \arcLoadMod_{[a]}^\beta(p') \Big]^2 \\
    &\hspace{1cm} \cdot \Bigg[ \frac{d\latency_{[a]}}{dw}\big( \bar \arcLoadMod_{[a]}^\beta(\bar \toll) \big) - \frac{d\latency_{[a]}}{dw}\big( \bar \arcLoadMod_{[a]}^\beta(p') \big) \Bigg] \\
    > \hspace{0.5mm} &0,
\end{align*}
which contradicts Theorem \ref{Thm: CoDAG is unique minimizer of F} .

The above arguments establish that if $\toll' \in \R^{|A_O|}$ satisfies \eqref{Eqn: Toll, Fixed Point Equation}, then $\bar \arcLoadMod_{[a]}^\beta(p') = \bar \arcLoadMod_{[a]}^\beta(\bar \toll)$ for each $[a] \in A_O$. Through \eqref{Eqn: Toll, Fixed Point Equation}, we then have, for each $[a] \in A_O$:
\begin{align*}
    \bar \toll_{[a]} &= \bar \arcLoadMod^\beta_{[a]}(\bar \toll) \cdot \frac{d\latency_{[a]}}{d w} \big(\bar \arcLoadMod^\beta_{[a]}(\bar \toll) \big) \\
    &= \bar \arcLoadMod^\beta_{[a]}(p') \cdot \frac{d\latency_{[a]}}{d w} \big(\bar \arcLoadMod^\beta_{[a]}(p') \big) \\
    &= \toll_{[a]}',
\end{align*}
so $\toll' = \bar \toll$. This concludes the proof.

\subsubsection{Proof of Lemma \ref{Lemma: p bar is Perturbed Socially Optimal}}
\label{subsubsec: A1, Proof of Lemma: p bar is Perturbed Socially Optimal}
In this subsection, we show that $\bar w^\beta(\bar \toll)$ is perturbed socially optimal.
Let $\arcLoadMod^\star \in \R^{|A|}$ denote the perturbed socially optimal load. Recall that, by Theorem \ref{Thm: CoDAG is unique minimizer of F} and the definition of the perturbed socially optimal load:
{\small
\begin{align*}
    &\bar \arcLoadMod^\beta(\bar \toll) \\
    = \hspace{0.5mm} &\text{arg}\min_{w \in \arcsLoadConstraintSet} \Bigg\{ \sum_{[a] \in A_O} \int_0^{\arcLoadMod_{[a]}} \big[ \latency_{[a]}(\costToGo) + \toll_{[a]} \big] dz \\
    &+ \frac{1}{\beta} \sum_{i \ne d} \Bigg[ \sum_{a \in A_i^+} \arcLoadMod_a \ln \arcLoadMod_a - \Bigg(\sum_{a \in A_i^+} \arcLoadMod_a \Bigg) \ln \Bigg(\sum_{a \in A_i^+} \arcLoadMod_a \Bigg) \Bigg] \Bigg\}, \\
    &w^\star \\
    = \hspace{0.5mm} &\text{arg}\min_{w \in \arcsLoadConstraintSet} \Bigg\{ \sum_{[a] \in A_O} \arcLoadMod_{[a]} \ln \arcLoadMod_{[a]} \\
    &+ \frac{1}{\beta} \sum_{i \ne d} \Bigg[ \sum_{a \in A_i^+} \arcLoadMod_a \ln \arcLoadMod_a - \Bigg(\sum_{a \in A_i^+} \arcLoadMod_a \Bigg) \ln \Bigg(\sum_{a \in A_i^+} \arcLoadMod_a \Bigg) \Bigg] \Bigg\}.
\end{align*}
}
The proof follows by verifying that the variational inequalities corresponding to the above two optimization problems are the same. These two variational inequalities in question are respectively given by:
\begin{align*}
    &\sum_{[a] \in A_O} \Bigg[ \latency_{[a]}\big(\bar \arcLoadMod_{[a]}^\beta(\bar \toll) \big) + \bar \arcLoadMod_{[a]}^\beta(\bar \toll) \frac{d\latency_{[a]}}{dw}\big( \bar \arcLoadMod_{[a]}^\beta(\bar \toll) \big) \\
    &\hspace{1cm} + \frac{1}{\beta} \ln\Bigg( \frac{\bar \arcLoadMod_{[a]}^\beta(\bar \toll)}{\sum_{a' \in A_{i_a}^+} \bar \arcLoadMod_{[a']}^\beta(\bar \toll)} \Bigg) \Bigg] \big( \arcLoadMod_a - \bar \arcLoadMod_{[a]}^\beta(\bar \toll) \big) > 0, \\
    &\hspace{1cm} \forall \hspace{0.5mm} w \in \arcsLoadConstraintSet, w \ne \bar \arcLoadMod_{[a]}^\beta(\bar \toll), \\
    &\sum_{[a] \in A_O} \Bigg[ \latency_{[a]}\big(\arcLoadMod_{[a]}^\star \big) + \arcLoadMod_{[a]}^\star \frac{d\latency_{[a]}}{dw}\big( \arcLoadMod_{[a]}^\star \big) \\
    &\hspace{1cm} + \frac{1}{\beta} \ln\Bigg( \frac{\bar \arcLoadMod_{[a]}^\star}{\sum_{a' \in A_{i_a}^+} \arcLoadMod_{[a']}^\star} \Bigg) \Bigg] \big( \arcLoadMod_a - \arcLoadMod_a^\star \big) > 0, \\
    &\hspace{1cm} \forall \hspace{0.5mm} w \in \arcsLoadConstraintSet, w \ne w^\star,
\end{align*}
and are thus, indeed, identical. This confirms that $\bar \arcLoadMod^\beta(\bar \toll) = w^\star$, and concludes the proof.

\subsection{Proofs for Section \ref{sec: Dynamics, Convergence}}
\label{subsec: A2, Dynamics Convergence}

\subsubsection{Statement of Lemma \ref{Lemma: Convergence, w, continuous}}
\label{subsubsec: A2, Statement of Lemma: Convergence, w, continuous}

% \frank{To edit, below:}

The complete, rigorous statement of Lemma 2, is as follows---Suppose $\arcLoadMod(0) \in \arcsLoadConstraintSet$, i.e., the initial flow satisfies flow continuity, and:
\begin{align*}
    K_i > \frac{g_o}{C_w} \max\{K_{i_{\hat a}}: \hat a \in \arcsMod_i^-\}
\end{align*}
for each $i \in \nodesMod \backslash \{d\}$, with $C_w$ given by Lemma \ref{Lemma: Technical Conditions for Stochastic Approximation, 1}. Then under the continuous-time flow dynamics \eqref{Eqn: General Network, w flow, continuous} and \eqref{Eqn: General Network, xi flow, continuous}, the continuous-time traffic allocation $\arcLoadMod(t)$ globally asymptotically converges to the corresponding CoDAG equilibrium $\bar \arcLoadMod^\beta(\toll)$.

The proof of Lemma \ref{Lemma: Convergence, w, continuous} follows by applying the proof of the analogous theorem in  \cite{Chiu2023ArcbasedTrafficAssignment} (Appendix C.1), and replacing the latencies $\latency_{[a]}(\arcLoadMod_{[a]})$ with the total cost $\latency_{[a]}(\arcLoadMod_{[a]}) + \toll_{[a]}$.

% Suppose $\arcLoadMod(0) \in \arcsLoadConstraintSet$, and:
% \begin{align*}
%     K_i \geq \frac{g_o}{C_w} \max\{K_{i_{\hat a}}: \hat a \in \arcsMod_i^-\}
% \end{align*}
% for each $i \in \nodesMod \backslash \{d\}$, with $C_w$ given by Lemma \ref{Lemma: Technical Conditions for Stochastic Approximation}. Then, the continuous-time dynamical system \eqref{Eqn: w flow, recursive, with h} for the traffic flow $\arcLoadMod(t)$ globally asymptotically converges to the corresponding Condensed DAG Equilibrium $\bar \arcLoadMod^\beta \in \arcsLoadConstraintSet$.

\subsubsection{Proof of Lemma \ref{Lemma: Technical Conditions for Stochastic Approximation, 1}}
\label{subsubsec: A2, Proof of Lemma: Technical Conditions for Stochastic Approximation, 1}

First, we rewrite the discrete $\probDist$-dynamics \eqref{Eqn: General Network, xi flow, discrete} as a Markov process with a martingale difference term:
\begin{align*}
    \probDist_a[n+1] &= \probDist_a[n] + \mu \big(\rho_a(\probDist[n], \tollDiscrete[n]) + M_a[n+1] \big),
\end{align*}
where $\rho_a: \R^{|\arcsMod|} \times \R^{|\arcsOrig|} \ra \R^{|\arcsMod|}$ is given by:
\begin{align} \label{Eqn: rho a}
    \rho_a(\probDist, \toll) &:= K_{i_a} \Bigg( -\probDist_a + \frac{\exp(-\beta \cdot  \costToGo_a(\arcLoadMod, \toll))}{\sum_{ a' \in \arcsMod_{i_a}^+} \exp(-\beta \cdot \costToGo_{a'}(\arcLoadMod, \toll))} \Bigg),
\end{align}
with $\arcLoadMod \in \R^{|\arcsMod|}$ defined arc-wise by $\arcLoadMod_a = (g_{i_a} + \sum_{\hat a \in A_{i_a}^-} w_{a'}) \cdot \probDist_a$, and:
\begin{align} \label{Eqn: Ma, Discrete-Time Dynamics}
    M_a[n+1] := \hspace{0.5mm} &\left( \frac{1}{\mu} \eta_{i_a}[n+1] - 1 \right) \cdot \rho_a(\probDist[n], \tollDiscrete[n]).
\end{align}
Here, $\arcLoadModDiscrete_a[n] = \big( \nodeLoadIn_{i_a} + \sum_{a' \in \arcsMod_{i_a}^-} W_{a'}[n] \big)$, as given by \eqref{Eqn: General Network, w flow, discrete}.

Below, we state and prove Lemma \ref{Lemma: Technical Conditions for Stochastic Approximation, 1}.

\begin{proof}
\begin{enumerate}
    \item We have:
    \begin{align*}
        &\E[M_a[n+1] | \mathcal{F}_n] \\
        = \hspace{0.5mm} &\left( \frac{1}{\mu} \E[\eta_{i_a}[n+1]] - 1 \right) \cdot K_{i_a} \\
        &\cdot \left(-\probDist_a[n] + \frac{\exp(-\beta \big[ \costToGo_a(\arcLoadModDiscrete[n], \tollDiscrete[n]) \big])}{\sum_{ a' \in \arcsMod_{i_a}^+} \exp(-\beta \big[ \costToGo_{a'}(\arcLoadModDiscrete[n], \tollDiscrete[n]) \big])} \right) \\
        = \hspace{0.5mm} &0.
    \end{align*}
    
    \item We separate the proof of this part of the lemma into the following steps. 
    \begin{itemize}
        \item First, we show that for each $a \in \arcsMod$, $n \geq 0$, we have $\probDist_a[n] \in (0, 1]$.
        
        $\hspace{5mm}$ Fix $a \in \arcsMod$ arbitrarily. Then $\probDist_a[0] \in (0, 1]$ by assumption, and for each $n \geq 0$:
        \begin{align*}
            \frac{\exp(-\beta \big[ \costToGo_a(\arcLoadModDiscrete[n], \tollDiscrete[n]) \big])}{\sum_{ a' \in \arcsMod_{i_a}^+} \exp(-\beta \big[ \costToGo_{a'}(\arcLoadModDiscrete[n], \tollDiscrete[n]) \big])} \in (0, 1],
        \end{align*}
        since the exponential function takes values in $(0, \infty)$. Thus, by Lemma \ref{Lemma: Technical Conditions for Stochastic Approximation, 1}, we have $\probDist_a[n] \in (0, 1]$ for each $n \geq 0$.
        
        \item Second, we show that for each $a \in \arcsMod$, $n \geq 0$, we have $\arcLoadModDiscrete_a[n] \in (0, g_o]$.
        
        $\hspace{5mm}$ Note that \eqref{Eqn: General Network, w flow, discrete}, together with the assumption that $\arcLoadModDiscrete[0] \in \arcsLoadConstraintSet$, implies that $\arcLoadModDiscrete[n] \in \arcsLoadConstraintSet$ for each $n \geq 0$. Now, fix $a \in \arcsMod$, $n \geq 0$ arbitrarily. Let $\routes(a) \subseteq \routes$ denote the set of all routes passing through $a$, and for each $r \in \routes(a)$, let $a_{r, k}$ denote the $k$-th arc in $r$. Then, by the conservation of flow encoded in $R$:
        \begin{align*}
            \arcLoadModDiscrete_a[n] &= g_o \cdot \sum_{r \in \routes(a)} \prod_{k=1}^{|r|} \probDist_{a_{r,k}} \\
            &\leq g_o \cdot \sum_{r \in \routes} \prod_{k=1}^{|r|} \probDist_{a_{r,k}} \\
            &= g_o.
        \end{align*}
        Similarly, since $\probDist_a[n] \in (0, 1]$ for each $a \in \arcsMod$, $n \geq 0$, we have:
        \begin{align*}
            \arcLoadModDiscrete_a[n] &= g_o \cdot \sum_{r \in \routes(a)} \prod_{k=1}^{|r|} \probDist_{a_{r,k}} > 0.
        \end{align*}
        
        \item Third, we show that there exists $C_p > 0$ such that $\tollDiscrete_a[n] \in [0, C_p]$ for each $a \in \arcsMod$, $n \geq 0$.
        
        $\hspace{5mm}$ Above, we have established that $\arcLoadModDiscrete_a[n] \in (0, g_o]$ for each $a \in \arcsMod$, $n \geq 0$. Moreover, by assumption, $\latency_{[a]}(\cdot)$ is non-negative, continuously differentiable, strictly increasing, and strictly convex. Thus, taking $C_{ds} := (d\latency_{[a]}/d\arcLoadMod)(g_o)$, we obtain:
        \begin{align*}
            \frac{d\latency_{[a]}}{d\arcLoadMod}(\arcLoadModDiscrete_{[a]}[n]) \in [0, C_{ds}].
        \end{align*}
        Now, take $C_p := \max\{\max_{a \in \arcsMod} \tollDiscrete_a[0], g_o C_{ds} \}$. By the definition of $C_p$, we have $\tollDiscrete_a[0] \in [0, C_p]$ for each $a \in \arcsMod$. Moreover, for each $n \geq 0$:
        \begin{align*}
            \arcLoadModDiscrete_{[a]}[n] \cdot \frac{d\latency_{[a]}}{d\arcLoadMod}\big(\arcLoadModDiscrete_{[a]}[n] \big) \in [0, g_o C_{ds}] \subseteq [0, C_p].
        \end{align*}
        Thus, by Lemma \ref{Lemma: Technical Conditions for Stochastic Approximation, 1}, we conclude that $\tollDiscrete_a[n] \in [0, C_p]$ for each $n \geq 0$.
        
        \item Fourth, we show that there exists $C_z > 0$ such that $|\costToGo_a(\arcLoadModDiscrete[n], \tollDiscrete[n])| \leq C_z$ for each $a \in \arcsMod$, $n \geq 0$. Fix $a \in \arcsMod_d^- = \{a \in \arcsMod: m_a = 1\}$ arbitrarily. Then, from \eqref{Eqn: CostToGo}:
        \begin{align*}
            &\costToGo_a(\arcLoadMod, \toll) = \latency_{[a]}(\arcLoadMod_{[a]}) + \toll_{[a]} \in [0, \latency_{[a]}(g_o) + C_p], \\
            \Ra \hspace{0.5mm} &|\costToGo_a(\arcLoadMod, \toll)| \leq \latency_{[a]}(g_o) + C_p := C_{z,1}.
        \end{align*}
        Now, suppose that at some height $k \in [\height(\graphMod) - 1]$, there exists some $C_{z,k} > 0$ such that, for each $n \geq 0$, and each $a \in \arcsMod$ satisfying $\height_a \leq k$ and each $n \geq 0$, we have $|\costToGo_a(\arcLoadMod, \toll)| \leq C_{z,k}$. Then, for each $n \geq 0$, and each $a \in \arcsMod$ satisfying $\height_a = k+1$ (at least one such $a \in \arcsMod$ must exist, by \cite{Chiu2023ArcbasedTrafficAssignment}, Proposition 2):
        \begin{align*}
            &\hspace{5mm} \costToGo_a(\arcLoadMod, \toll) \\
            &= \latency_{[a]}(\arcLoadMod_{[a]}) + \toll_{[a]} - \frac{1}{\beta} \ln\left( \sum_{a' \in \arcsMod_{j_a}^+} e^{-\beta \cdot \costToGo_{a'}(\arcLoadMod, \toll)} \right) \\
            &\leq \latency_{[a]}(g_o) + C_p - \frac{1}{\beta} \ln\left( |\arcsMod_{j_a}^+| e^{-\beta \cdot C_z} \right) \\
            &= \latency_{[a]}(g_o) + C_p + C_z,
        \end{align*}
        and:
        \begin{align*}
            &\hspace{5mm} \costToGo_a(\arcLoadMod, \toll) \\
            &= \latency_{[a]}(\arcLoadMod_{[a]}) + \toll_{[a]} - \frac{1}{\beta} \ln\left( \sum_{a' \in \arcsMod_{j_a}^+} e^{-\beta \cdot \costToGo_{a'}(\arcLoadMod, \toll)} \right) \\
            &\geq 0 + 0 - \frac{1}{\beta} \ln\left( |\arcsMod_{j_a}^+| e^{\beta \cdot C_z} \right) \\
            &= - \frac{1}{\beta}\ln|\arcsMod| - C_z ,
        \end{align*}
        from which we conclude that:
        \begin{align*}
            &\hspace{5mm} |\costToGo_a(\arcLoadMod, \toll)| \\
            &\leq \max\left\{\latency_{[a]}(g_o) + C_p + C_z, \frac{1}{\beta}\ln|\arcsMod| + C_z \right\} \\
            &:= C_{z,k+1},
        \end{align*}
        with $C_{z+1} \geq C_z$. This completes the induction step, and the proof is completed by taking $C_z := C_{z, \height(\graphMod)}$.
        
        \item Fifth, we show that there exists some $C_\probDist > 0$ such that $\probDist_a[n] \geq C_\probDist$ for each $a \in \arcsMod$, $n \geq 0$. 
        
        $\hspace{5mm}$ Define:
        \begin{align*}
            C_\probDist := \min\left\{ \min\{\probDist_{a'}[0]: a' \in \arcsMod\}, \frac{1}{|\arcsMod|} e^{-2 \beta C_z} \right\} > 0.
        \end{align*}
        By definition of $C_\probDist$, we have $\probDist_a[0] \geq C_\probDist$. Moreover, for each $n \geq 0$, we have:
        \begin{align*}
            &\frac{\exp(-\beta \big[ \costToGo_a(\arcLoadModDiscrete[n], \tollDiscrete[n]) \big])}{\sum_{ a' \in \arcsMod_{i_a}^+} \exp(-\beta \big[ \costToGo_{a'}(\arcLoadModDiscrete[n], \tollDiscrete[n]) \big])} \\
            \geq \hspace{0.5mm} &\frac{e^{-\beta C_z}}{|\arcsMod_{i_a}^+| \cdot e^{\beta C_z}} \\
            \geq \hspace{0.5mm} & \frac{1}{|\arcsMod|} e^{-2\beta C_z} \\
            \geq \hspace{0.5mm} &C_\probDist.
        \end{align*}
        Thus, by Lemma \ref{Lemma: Technical Conditions for Stochastic Approximation, 1}, we have $\probDist_a[n] \geq C_\probDist$ for each $n \geq 0$.
        
        \item Sixth, we show that there exists $C_w > 0$ such that, for each $a \in \arcsMod$, $n \geq 0$, we have $\arcLoadModDiscrete_a[n] \geq C_w$.
        
        $\hspace{5mm}$ Fix $a \in \arcsMod$, $n \geq 0$. Let $r \in \routes$ be any route in the corresponding DAG containing $a \in \arcsMod$. By unwinding the recursive definition of $\arcLoadModDiscrete_a[n]$ from the flow allocation probability values $\{\probDist_a[n]: a \in \arcsMod, n \geq 0\}$, we have:
        \begin{align*}
            \arcLoadModDiscrete_a[n] &= g_o \cdot \sum_{\substack{r' \in \routes \\ a \in r'}} \prod_{a' \in r'} \probDist_{a'}[n] \\
            &\geq g_o \cdot \prod_{a' \in r} \probDist_{a'}[n] \\
            &\geq g_o \cdot (C_\probDist)^{|r|} \\
            &\geq g_o \cdot (C_\probDist)^{\depth(\graphMod)} \\
            &:= C_w.
        \end{align*}
        
        \item Seventh, we show that there exists $C_m > 0$ such that, for each $a \in \arcsMod$, $n \geq 0$, we have $M_a[n] \geq C_m$.
        
        $\hspace{5mm}$ Define, for convenience, $C_\mu := \max\{\overline \mu - \mu, \mu - \underline \mu\}$. Since $\eta_{i_a}[n] \in [\underline \mu, \overline \mu]$, we have from \eqref{Eqn: Ma, Discrete-Time Dynamics} that for each $a \in \arcsMod$, $n \geq 0$:
        {\small 
        \begin{align*}
            &M_a[n+1] \\
            = \hspace{0.5mm} &\left( \frac{1}{\mu} \eta_{i_a}[n+1] - 1 \right) \cdot K_{i_a} \\
            &\cdot \left(- \probDist_a[n] + \frac{\exp(-\beta \big[ \costToGo_a(\arcLoadModDiscrete[n], \tollDiscrete[n]) \big])}{\sum_{ a' \in \arcsMod_{i_a}^+} \exp(-\beta \big[ \costToGo_{a'}(\arcLoadModDiscrete[n], \tollDiscrete[n]) \big])} \right).
        \end{align*}
        }
        Applying the triangle inequality, we obtain:
        \begin{align*}
            |M_a[n+1]| &\leq \frac{1}{\mu} K_{i_a} C_\mu \cdot (1+1) \\
            &= \frac{2}{\mu} C_\mu \cdot \max_{i \in \nodesMod \backslash \{d\}} K_i \\
            &:= C_m.
        \end{align*}
    \end{itemize}
    
    \item We separate the proof of this part of the lemma into the following steps.
    
    \begin{itemize}
        \item First, we show that for each $a \in \arcsMod$, $t \geq 0$, we have $\probDist_a(t) \in (0, 1]$.
        
        $\hspace{5mm}$ Fix $a \in \arcsMod$. By assumption, $\probDist_a(0) \in (0, 1]$, and at each $t \geq 0$:
        \begin{align*}
            \frac{\exp(-\beta \costToGo_a(\arcLoadMod, \toll))}{\sum_{a' \in \arcsMod_{i_a}^+} \exp(-\beta \costToGo_{a'}(\arcLoadMod, \toll))} \in (0, 1].
        \end{align*}
        Thus, by Lemma \ref{Lemma: Technical Conditions for Stochastic Approximation, 1}, we conclude that $\probDist_a(t) \in (0, 1]$ for each $t \geq 0$.
        
        % $\hspace{5mm}$ Next, recall from Theorem \ref{Thm: Convergence, w} that $\arcLoadMod(t) \in \arcsLoadConstraintSet$ for each $t \geq 0$. From the third and fourth bullet points in the second part of this Proposition, we find that $\tollDiscrete_a[]$
        
        \item Second, we show that $\arcLoadMod_a(t) \in [0, g_o]$ for each $t \geq 0$.
        
        $\hspace{5mm}$ The proof here is nearly identical to the proof that $\arcLoadModDiscrete_a[n] \in (0, g_o)$ in the second bullet point of the second part of this Proposition, and is omitted for brevity.
        
        \item Third, we show that $\toll_a(t) \in [0, C_p]$ for each $t \geq 0$. 
        
        $\hspace{5mm}$ Above, we have established that $\arcLoadMod_a(t) \in (0, g_o]$ for each $a \in \arcsMod$, $t \geq 0$. Let $C_p > 0$ be as defined in the third bullet point of the second part of this Proposition, i.e., $C_p = \max\left\{\max_{a \in \arcsMod} \tollDiscrete_a[0], g_o \cdot \frac{d\latency_{[a]}}{d\arcLoadMod}(g_o) \right\}$. Then, for each $a \in \arcsMod$, $t \geq 0$, we have:
        \begin{align*}
            \arcLoadMod_a(t) \cdot \frac{d\latency_{[a]}}{d\arcLoadMod}(\arcLoadMod_a(t)) \in [0, C_p].
        \end{align*}
        Note also that $\tollDiscrete_a[0] \leq C_p$ for each $a \in \arcsMod$, by definition of $C_p$. Thus, Lemma \ref{Lemma: Technical Conditions for Stochastic Approximation, 1} implies that $\toll_a(t) \in [0, C_p]$ for each $t \geq 0$.
        
        \item Fourth, we show that $|\costToGo_a(\arcLoadMod_a(t), \toll_a(t))| \leq C_z$ for each $t \geq 0$.
        
        $\hspace{5mm}$ The proof here is nearly identical to the proof that $|\costToGo_a(\arcLoadModDiscrete_a[n], \tollDiscrete_a[n])| \leq C_z$ in the fourth bullet point of the second part of this Proposition, and is omitted for brevity.
        
        \item Fifth, we show that there exists some $C_\probDist > 0$ such that $\probDist_a(t) \geq C_\probDist$ for each $a \in \arcsMod$, $t \geq 0$.  
        
        $\hspace{5mm}$ Define:
        \begin{align*}
            C_\probDist := \min\left\{ \min\{\probDist_{a'}(0): a' \in \arcsMod\}, \frac{1}{|\arcsMod|} e^{-2 \beta C_z} \right\}
            > 0.
        \end{align*}
        By definition of $C_\probDist$, we have $\probDist_a(0) \geq C_\probDist$. Moreover, for each $n \geq 0$, we have:
        \begin{align*}
            &\hspace{5mm} \frac{\exp(-\beta \big[ \costToGo_a(\arcLoadModDiscrete[n], \tollDiscrete[n]) \big])}{\sum_{ a' \in \arcsMod_{i_a}^+} \exp(-\beta \big[ \costToGo_{a'}(\arcLoadModDiscrete[n], \tollDiscrete[n]) \big])} \\
            &\geq \frac{e^{-\beta C_z}}{|\arcsMod_{i_a}^+| \cdot e^{\beta C_z}} \\
            &\geq \frac{1}{|\arcsMod|} e^{-2\beta C_z} \\
            &\geq C_\probDist.
        \end{align*}
        Thus, by Lemma \ref{Lemma: Technical Conditions for Stochastic Approximation, 1}, we have $\probDist_a(t) \geq C_\probDist$ for each $t \geq 0$.
 
        \item Sixth, we show that there exists $C_w > 0$ such that, for each $a \in \arcsMod$, $t \geq 0$, we have $\arcLoadMod_a(t) \geq C_w$.
        
        $\hspace{5mm}$ The proof here is nearly identical to the proof that $\arcLoadModDiscrete_a[n] \geq C_w$ in the fourth bullet point of the second part of this Proposition, and is omitted for brevity.
        
    \end{itemize}
    
\end{enumerate}

\end{proof}

Below, we state and prove Lemma \ref{Lemma: Technical Conditions for Stochastic Approximation, 2}, which together with Lemma \ref{subsubsec: A1, Proof of Lemma: C1 of CoDAG Equilibrium} supplies all the technical conditions necessary for Borkar's stochastic approximation theory to be applied.

\subsubsection{Proof of Lemma \ref{Lemma: Technical Conditions for Stochastic Approximation, 2}}
\label{subsubsec: A2, Proof of Lemma: Technical Conditions for Stochastic Approximation, 2}

~\\
\begin{proof}
\begin{enumerate} 
    \item Since $\bar \probDist^\beta(\toll)$ can be derived component-wise from $\bar \arcLoadMod^\beta$, we first show that $\bar \arcLoadMod^\beta: \R^{|\arcsOrig|} \ra \R^{|\arcsMod|}$ is Lipschitz continuous. We do so by showing that $\arcLoadMod^\beta$ is continuously differentiable with bounded derivative. To this end, recall from the proofs of Lemma \ref{Lemma: C1 of CoDAG Equilibrium}  and Lemma \ref{Lemma: Convergence, p, continuous}, the matrix $M \in \R^{|\arcsMod| \times d}$, $b \in \R^d$, with $d \in [|\arcsMod|]$ describing the dimension of $\arcsLoadConstraintSet$ (as a manifold with boundary), and the matrices $B \in \R^{|\arcsMod| \times (|\arcsMod| - d)}$, $C \in \R^{|\arcsMod| \times |\arcsOrig|}$.
    
    $\hspace{5mm}$ As established in the proof of Proposition \ref{Lemma: C1 of CoDAG Equilibrium}, there exists a continuously differentiable function $J: \R^{|\arcsOrig|} \times \R^{|\arcsMod|} \ra \R^{|\arcsMod|}$, and matrices $M \in \R^{|\arcsMod| \times d}$ and $B \in \R^{|\arcsMod| \times (|\arcsMod| - d)}$, such that $J(p, \arcLoadMod^\beta(\toll)) = 0$ for each $\toll \in \R^{|\arcsOrig|}$, the columns of $B$ and the columns of $M$ are orthonormal, $R(M)$ and $R(B)$ are orthogonal subspaces whose direct sum is $\R^{|\arcsMod|}$, and:
\begin{align*}
    \frac{\partial J}{\partial \arcLoadMod}(\arcLoadMod, \toll) &= \begin{bmatrix}
        M^\top \\
        B^\top \nabla_\arcLoadMod^2 F(\arcLoadMod, \toll)
    \end{bmatrix} \in \R^{|\arcsMod| \times |\arcsMod|},
\end{align*}
where, as in the proof of Proposition \ref{Lemma: C1 of CoDAG Equilibrium}, $F: \arcsLoadConstraintSet \times \R^{|\arcsOrig|} \ra \R$ is given by \eqref{Eqn: Def, F}, reproduced below:
\begin{align} \nonumber
    &F(\arcLoadMod, \toll) \\ \nonumber
    := \hspace{0.5mm} &\sum_{[a] \in A_0} \int_0^{\arcLoadMod_{[a]}} \big[ \latency_{[a]}(\costToGo) + \toll_{[a]} \big] \hspace{0.5mm} dz \\ \nonumber
    + &\frac{1}{\beta} \sum_{i \ne d} \Bigg[ \sum_{a \in A_i^+} \arcLoadMod_a \ln \arcLoadMod_a - \Bigg(\sum_{a \in A_i^+} \arcLoadMod_a \Bigg) \ln \Bigg(\sum_{a \in A_i^+} \arcLoadMod_a \Bigg) \Bigg].
\end{align}
Thus, the Implicit Function Theorem implies that:
\begin{align} \nonumber
    &\hspace{5mm} \frac{d \bar \arcLoadMod^\beta}{d\toll}(\toll) \\ \nonumber
    &= - \Bigg[ \frac{\partial J}{\partial \arcLoadMod}(\bar \arcLoadMod^\beta(\toll), \toll) \Bigg]^{-1} \frac{\partial J}{\partial \toll}(\bar \arcLoadMod^\beta(\toll), \toll) \\ \label{Eqn: d bar w beta d\toll Expression, from Implicit Function Theorem}
    &= - \begin{bmatrix}
        M^\top \\
        B^\top \nabla_\arcLoadMod^2 F(\bar \arcLoadMod^\beta(\toll), \toll)
    \end{bmatrix}^{-1} 
    \begin{bmatrix}
        O \\
        B^\top \frac{d}{d\toll} \nabla_\arcLoadMod F(\bar \arcLoadMod^\beta(\toll), \toll)
    \end{bmatrix},
\end{align}
where $\nabla_\arcLoadMod F(\arcLoadMod, \toll) \in \R^{|\arcsMod|}$, and $\frac{\partial}{\partial \toll} \nabla_\arcLoadMod F(\arcLoadMod, \toll) \in \R^{|\arcsMod| \times |\arcsOrig|}$. To study \eqref{Eqn: d bar w beta d\toll Expression, from Implicit Function Theorem} further, we wish to rewrite the $B^\top \nabla_\arcLoadMod^2 F(\arcLoadMod, \toll)$ term. To this end, note that since $\begin{bmatrix}
    M & B
\end{bmatrix} \in \R^{|\arcsMod| \times |\arcsMod|}$ is an orthogonal matrix, and $\nabla_\arcLoadMod^2 F(\arcLoadMod, \toll)$ is symmetric positive definite (since $F(p, \cdot)$ is strictly convex for each $\toll \in \R^{|\arcsOrig|}$), the matrix:
\begin{align*}
    Q := \begin{bmatrix}
        M^\top \\ B^\top
    \end{bmatrix} 
    \nabla_\arcLoadMod^2 F(\bar \arcLoadMod^\beta(\toll), \toll)
    \begin{bmatrix}
        M & B
    \end{bmatrix} \in \R^{|\arcsMod| \times |\arcsMod|}
\end{align*}
is symmetric positive definite as well. Now, let $Q_{11} := M^\top \nabla_\arcLoadMod^2 F(\arcLoadMod, \toll) M \in \R^{d \times d}, Q_{12} := M^\top \nabla_\arcLoadMod^2 F(\arcLoadMod, \toll) B \in \R^{d \times (|\arcsMod| - d)}$, and $Q_{22} := B^\top \nabla_\arcLoadMod^2 F(\arcLoadMod, \toll) B \in \R^{(|\arcsMod| - d) \times (|\arcsMod| - d)}$ denote the various block matrices of $Q$, as shown below:
\begin{align*}
    Q = \begin{bmatrix}
        Q_{11} & Q_{12} \\
        Q_{12}^\top & Q_{22}
    \end{bmatrix}.
\end{align*}
We then have:
\begin{align*}
    B^\top \nabla_\arcLoadMod^2 F(\arcLoadMod, \toll) &= \begin{bmatrix}
        O & I
    \end{bmatrix}
    \begin{bmatrix}
        M^\top \\ B^\top
    \end{bmatrix} \nabla_\arcLoadMod^2 F(\arcLoadMod, \toll) \\
    &= \begin{bmatrix}
        O & I
    \end{bmatrix} Q
    \begin{bmatrix}
        M^\top \\ B^\top
    \end{bmatrix} \\
    &= Q_{12}^\top M^\top + Q_{22} B^\top,
\end{align*}
where the matrices $O$ and $i \in \nodesMod$ above are the zero matrix of dimension $(|\arcsMod| - d) \times d$ and identity matrix of dimension $(|\arcsMod| - d) \times (|\arcsMod| - d)$, respectively. Substituting back into \eqref{Eqn: d bar w beta d\toll Expression, from Implicit Function Theorem}, we obtain:
\begin{align} \nonumber
    &\hspace{5mm}\frac{d \bar \arcLoadMod^\beta}{d\toll}(\toll) \\
    &= - \begin{bmatrix}
        M^\top \\
        B^\top \nabla_\arcLoadMod^2 F(\arcLoadMod, \toll)
    \end{bmatrix}^{-1} 
    \begin{bmatrix}
        O \\
        B^\top \frac{d}{d\toll} \nabla_\arcLoadMod F(\arcLoadMod, \toll)
    \end{bmatrix} \\ \nonumber
    &= - \begin{bmatrix}
        M^\top \\
        Q_{12}^\top M^\top + Q_{22} B^\top
    \end{bmatrix}^{-1} 
    \begin{bmatrix}
        O \\
        B^\top \frac{d}{d\toll} \nabla_\arcLoadMod F(\arcLoadMod, \toll)
    \end{bmatrix} \\ \nonumber
    &= - \Bigg( \begin{bmatrix}
        I & O \\
        Q_{12}^\top & Q_{22}
    \end{bmatrix} 
    \begin{bmatrix}
        M^\top \\ B^\top
    \end{bmatrix}
    \Bigg)^{-1} 
    \begin{bmatrix}
        O \\
        B^\top 
    \end{bmatrix}
    \frac{d}{d\toll} \nabla_\arcLoadMod F(\arcLoadMod, \toll) \\ \label{Eqn: d bar w beta d\toll, Final Expression, 1}
    &= - \begin{bmatrix}
        M & B
    \end{bmatrix} \begin{bmatrix}
        I & O \\
        - Q_{22}^{-1} Q_{12}^\top & Q_{22}^{-1}
    \end{bmatrix} 
    \begin{bmatrix}
        O \\
        B^\top 
    \end{bmatrix}
    \frac{d}{d\toll} \nabla_\arcLoadMod F(\arcLoadMod, \toll) \\ \label{Eqn: d bar w beta d\toll, Final Expression, 2}
    &= -B Q_{22}^{-1} B^\top C \\ \nonumber
    &= -B (B^\top \nabla_\arcLoadMod^2 F(\arcLoadMod, \toll) B)^{-1} B^\top C.
\end{align}
    
    % Recall also that, in Lemma \ref{Lemma: Convergence, p, continuous}, the derivative of $\bar \arcLoadMod^\beta$ is given by \eqref{Eqn: d bar w beta dp, Final Expression, 2}, reproduced below:
    % \begin{align*}
    %     \frac{d\bar \arcLoadMod^\beta}{d\toll}(\toll) &= - B (B^\top \nabla_w^2 F(\bar \arcLoadMod^\beta(\toll), \toll) B)^{-1} B^\top C,
    % \end{align*}
    % for each $\toll \in \R^{|\arcsOrig|}$. 
    Below, to establish the Lipschitz continuity of $\bar \arcLoadMod^\beta(\cdot)$, we provide a uniform bound for $\frac{d\bar \arcLoadMod^\beta}{d\toll}(\toll)$ over all values of $\toll \in \R^{|\arcsOrig|}$, by providing a uniform upper bound for the minimum eigenvalue of $\nabla_w^2 F(\bar \arcLoadMod^\beta(\toll)$ over all values of $\toll \in \R^{|\arcsOrig|}$.
    
    $\hspace{5mm}$ From Lemma \ref{Lemma: Technical Conditions for Stochastic Approximation, 1}, $\bar \arcLoadMod_a^\beta(\toll) \in [C_w, g_o]$ for each $\toll \in \R^{|\arcsOrig|}$ and $a \in \arcsMod$. Thus, all the second partial derivatives of $F$, as given by:
    \begin{align*}
        &\hspace{5mm} \frac{\partial^2}{\partial \arcLoadMod_a \partial \arcLoadMod_{a'}} F(\bar \arcLoadMod^\beta(\toll), \toll) \\
        &= \left(\frac{d\latency_{[a]}}{d\arcLoadMod}(\bar \arcLoadMod^\beta(\toll)) + \frac{1}{\bar \arcLoadMod_a^\beta(\toll)} \right) \cdot \textbf{1}\{a = a'\} \\
        &\hspace{1.5cm} - \frac{1}{\sum_{a' \in \arcsMod_{i_a}^+} \bar \arcLoadMod_{a'}^\beta(\toll)} \cdot \textbf{1}\{i_{a'} = i_a\}
    \end{align*}
    are well-defined and continuous. Next, consider the continuous map from each $\toll \in \R^|\arcsOrig|$ to the minimum eigenvalue of $\nabla_w^2 F(\bar \arcLoadMod^\beta(\toll), \toll)$, officially stated as:
    \begin{align*}
        p \mapsto \min_{\substack{v \in \R^{|\arcsMod|} \\ \Vert v \Vert_2 = 1}} v^\top \nabla_w^2 F(\bar \arcLoadMod^\beta(\toll), \toll) v.
    \end{align*}
    Since $F$ is strictly convex, $\nabla_w^2 F(\bar \arcLoadMod^\beta(\toll), \toll)$ is symmetric positive definite for each $\toll \in \R^{|\arcsOrig|}$, meaning that the output of the above map is strictly positive for each $\toll \in \R^{|\arcsOrig|}$. Moreover, note that the entries of $F(\overline \arcLoadMod^\beta(\toll), \toll)$ only depend on $\toll$ through the value of $\overline \arcLoadMod^\beta(\toll)$, which is bounded in $[C_w, g_o]^{|\arcsMod|}$. Thus, for each $\toll \in \R^{|\arcsOrig|}$:
    \begin{align*}
        &\hspace{5mm} \min_{\substack{v \in \R^{|\arcsMod|} \\ \Vert v \Vert_2 = 1}} v^\top \nabla_w^2 F(\bar \arcLoadMod^\beta(\toll), \toll) v \\
        &\geq \min_{w \in [C_w, g_o]^{|\arcsMod|}} v^\top \nabla_w^2 F(\arcLoadMod, \toll) v \\
        &:= C_F > 0,
    \end{align*}
    where $C_F := \min_{w \in [C_w, g_o]^{|\arcsMod|}} v^\top \nabla_w^2 F(\arcLoadMod, \toll) v$ is independent of $\toll$, and is strictly positive, since the minimum of a strictly positive-valued function over a compact set is strictly positive. We thus have a uniform bound on the derivative of $\bar \arcLoadMod^\beta$ over all values of $\toll \in \R^{|\arcsOrig|}$ at which it is evaluated:
    \begin{align*}
        &\hspace{5mm} \left\Vert \frac{d\bar \arcLoadMod^\beta}{d\toll}(\toll) \right\Vert_2 \\
        &\leq \Vert B \Vert_2^2 \Vert C \Vert_2 \cdot \Vert (B^\top \nabla_w^2 F(\bar \arcLoadMod^\beta(\toll), \toll) B)^{-1} \Vert_2 \\
        &= \Vert B \Vert_2^2 \Vert C \Vert_2 \cdot \frac{1}{\min\limits_{\substack{\hat v \in \R^{|\arcsMod|-d} \\ \Vert \hat v \Vert_2 = 1}} \hat v^\top B^\top \nabla_w^2 F(\bar \arcLoadMod^\beta(\toll), \toll) B \hat v} \\
        &\leq \Vert B \Vert_2^2 \Vert C \Vert_2 \cdot \frac{1}{\min\limits_{\substack{v \in \R^{|\arcsMod|} \\ \Vert v \Vert_2 = 1}} v^\top \nabla_w^2 F(\bar \arcLoadMod^\beta(\toll), \toll) v} \\
        &\leq \Vert B \Vert_2^2 \Vert C \Vert_2 \cdot \frac{1}{C_F}.
    \end{align*}
    
    \item We shall establish the Lipschitz continuity of (the restriction of) $\costToGo_a$, for each $a \in \arcsMod$, by providing uniform bounds on its partial derivatives across all values of its arguments $(\arcLoadMod, \toll) \in \arcsLoadConstraintSet' \times [0, C_p]^{|\arcsOrig|}$.
    
    $\hspace{5mm}$ The proof follows by induction on the height index $k \in [\height(\graphMod)]$. For each $a \in \arcsMod$, let $\tilde \costToGo_a: \R^{|\arcsMod|} \ra \R$ be the continuous extension of $\costToGo_a: \arcsLoadConstraintSet \ra \R$ to the Euclidean space $\R^{|\arcsMod|}$ containing $\arcsLoadConstraintSet$. By definition of Lipschitz continuity, if $\tilde \costToGo_a$ is Lipschitz for some $a \in \arcsMod$, then so is $\costToGo_a$. For each $a \in \arcsMod_d^- = \{a \in \arcsMod: \height_a = 1\}$ and any $\arcLoadMod \in \R^{|\arcsMod|}$:
    \begin{align*}
        \tilde \costToGo_a(\arcLoadMod) = \latency_{[a]}(\arcLoadMod_{[a]}) + \toll_{[a']}.
    \end{align*}
    Thus, for any $\hat a \in \arcsMod$, and any $\arcLoadMod \in \R^{|\arcsMod|}$, $\toll \in \R^{|\arcsOrig|}$:
    \begin{align*}
        \frac{\partial \tilde \costToGo_a}{\partial \arcLoadMod_{\hat a}}(\arcLoadMod, \toll) &= \frac{d\latency_{[a]}}{d\arcLoadMod}(\arcLoadMod_{[a]}) \cdot \textbf{1}\{\hat a \in [a]\} \in [0, C_{ds}], \\
        \frac{\partial \tilde \costToGo_a}{\partial \toll_{[\hat a}]}(\arcLoadMod, \toll) &= \textbf{1}\{\hat a \in [a]\} \in [0, 1].
    \end{align*}
    We set $C_{z,1} := \max\{C_{ds}, 1\}$.
    
    $\hspace{5mm}$ Now, suppose that there exists some depth $k \in [\height(\graphMod) - 1]$ and some constant $C_{z,k} > 0$ such that, for any $a \in \arcsMod$ satisfying $\height_a \leq k$, and any $\arcLoadMod \in \arcsLoadConstraintSet$, $n \geq 0$, the map $\tilde \costToGo_a: \R^{|\arcsMod|} \ra \R$ is continuously differentiable, with:
    \begin{align*}
        \left|\frac{\partial \tilde \costToGo_a}{\partial \arcLoadMod_{\hat a}}(\arcLoadMod) \right| &\leq C_{z,k}, \hspace{5mm} \left|\frac{\partial \tilde \costToGo_a}{\partial \toll_{[\hat a]}}(\arcLoadMod) \right| \leq C_{z,k}.
    \end{align*}
    Continuing with the induction step, fix $a \in \arcsMod$ such that $\height_a = k+1$ (there exists at least one such link, by \cite{Chiu2023ArcbasedTrafficAssignment}, Proposition 1, Part 4). From \cite{Chiu2023ArcbasedTrafficAssignment}, Proposition 1, Part 2, we have $\height_{a'} \leq k$ for each $a' \in \arcsMod_{i_a}^+$. Thus, the induction hypothesis implies that, for any $\hat a \in \arcsMod$:
    \begin{align*}
        \tilde \costToGo_a(\arcLoadMod, \toll) &= \latency_{[a]}(\arcLoadMod_{[a]}) + \toll_{[a]} - \frac{1}{\beta} \sum_{a' \in \arcsMod_{i_a}^+} e^{-\beta \costToGo_{a'}(\arcLoadMod, \toll)}.
    \end{align*}
    Computing partial derivatives with respect to each component of $\arcLoadMod$, we obtain:
    \begin{align*}
        \frac{\partial \tilde \costToGo_a}{\partial \arcLoadMod_{\hat a}}(\arcLoadMod, \toll) &= \frac{d\latency_{[a]}}{d\arcLoadMod}(\arcLoadMod_{[a]}) \cdot \textbf{1}\{\hat a \in [a]\} \\
        &\hspace{5mm} + \sum_{a' \in \arcsMod_{j_a}^+} e^{-\beta \tilde \costToGo_{a'}(\arcLoadMod, \toll)} \cdot \frac{\partial \tilde \costToGo_{a'}}{\partial \arcLoadMod_{\hat a}}(\arcLoadMod, \toll), \\
        \Ra \hspace{0.5mm} \left| \frac{\partial \tilde \costToGo_a}{\partial \arcLoadMod_{\hat a}}(\arcLoadMod) \right| &\leq C_{ds} + |\arcsMod| \cdot C_{z,k}.
    \end{align*}
    Computing partial derivatives with respect to each component of $\toll$, we obtain:
    \begin{align*}
        \frac{\partial \tilde \costToGo_a}{\partial \toll_{[\hat a]}}(\arcLoadMod, \toll) &= \textbf{1}\{\hat a \in [a]\} \\
        &\hspace{5mm} + \sum_{a' \in \arcsMod_{j_a}^+} e^{-\beta \tilde \costToGo_{a'}(\arcLoadMod, \toll)} \cdot \frac{\partial \tilde \costToGo_{a'}}{\partial \toll_{\hat a}}(\arcLoadMod, \toll), \\
        \Ra \hspace{0.5mm} \left| \frac{\partial \tilde \costToGo_a}{\partial \arcLoadMod_{\hat a}}(\arcLoadMod) \right| &\leq 1 + |\arcsMod| \cdot C_{z,k}.
    \end{align*}
    We can complete the induction step by taking $C_{z,k+1} := \max\{C_{ds}, 1\} + |\arcsMod| \cdot C_{z,k}$.
    
    $\hspace{5mm}$ This establishes that, for each $a \in \arcsMod$, the map $\costToGo_a$ is continuously differentiable, with partial derivatives uniformly bounded by a uniform constant, $C_z := C_{z,\height(\graphMod)}$. This establishes the Lipschitz continuity of the map $\costToGo_a$ for each $a \in \arcsMod$, and thus proves this part of the proposition.
    
    \item Recall that the map from traffic allocation probabilities ($\probDist$) to traffic flows ($\arcLoadMod$) is given as follows, for each $a \in \arcsMod$:
    \begin{align*}
        \arcLoadMod_a = \left(g_{i_a} + \sum_{\hat a \in \arcsMod_i^-} \arcLoadMod_a \right) \cdot \probDist_a = g_o \cdot \sum_{\substack{r \in \routes \\ a \in r}} \prod_{k=1}^{|r|} \probDist_{a_{r,k}},
    \end{align*}
    where $a_{r,k}$ denotes the $k$-th arc along a given route $r \in \routes$, for each $k \in |r|$. It is there clear that the map from $\probDist$ to $\arcLoadMod$ is continuously differentiable. Moreover, the domain of this map is compact; indeed, for each $a \in \arcsMod$, we have $\probDist_a \in [0, 1]$, and for each non-destination node $i \ne d$, we have $\sum_{a \in \arcsMod_i^+} \probDist_a = 1$. Thus, the map $\probDist \mapsto \arcLoadMod$ has continuously differentiable derivatives with magnitude bounded above by some constant uniform in the compact set of realizable probability allocations $\probDist$. This is equivalent to stating that the map $\probDist \mapsto \arcLoadMod$ is Lipschitz continuous.
    
    \item Above, we have established that the maps $\costToGo_a$ and $\probDist \mapsto \arcLoadMod$ are Lipschitz continuous. Since the addition and composition of Lipschitz maps is Lipschitz, it suffices to verify that the map $\hat \rho: \R^{|\arcsMod|} \ra \R^{|\arcsMod|}$, defined element-wise by:
    \begin{align*}
        \hat \rho_a(z) := \frac{e^{-\beta \costToGo_a}}{\sum_{a' \in \arcsMod_{i_a}^+} e^{-\beta \costToGo_{a'}}}, \hspace{1cm} \forall \hspace{0.5mm} a \in \arcsMod
    \end{align*}
    is Lipschitz continuous. We do so below by computing, and establishing a uniform bound for, its partial derivatives. For each $\hat a \in \arcsMod$:
    \begin{align*}
        \frac{\partial \hat \rho_a}{\partial \costToGo_{\bar a}} &= \frac{1}{(\sum_{a' \in \arcsMod_{i_a}^+} e^{-\beta \costToGo_{a'}})^2} \\
        &\hspace{1cm} \cdot \Bigg( \sum_{a' \in \arcsMod_{i_a}^+} e^{-\beta \costToGo_{a'}} \cdot (-\beta) e^{-\beta \costToGo_a} \cdot \frac{\partial \costToGo_a}{\partial \costToGo_{\bar a}} \\
        &\hspace{5mm} - e^{-\beta \costToGo_a} \cdot \sum_{a' \in \arcsMod_{i_a}^+} (-\beta) e^{-\beta \costToGo_{a'}} \frac{\partial \costToGo_{a'}}{\partial \costToGo_{\bar a}} \Bigg), \\
        &= \frac{e^{-\beta \costToGo_a}}{\sum_{a' \in \arcsMod_{i_a}^+} e^{-\beta \costToGo_{a'}}} \cdot \beta \cdot \frac{\partial \costToGo_a}{\partial \costToGo_{\hat a}} \\
        &\hspace{5mm} + \frac{\beta e^{-\beta \costToGo_a}}{(\sum_{a' \in \arcsMod_{i_a}^+} e^{-\beta \costToGo_{a'}})^2} \cdot \sum_{a' \in \arcsMod_{i_a}^+} e^{-\beta \costToGo_{a'}} \frac{\partial \costToGo_{a'}}{\partial \costToGo_{\bar a}},
    \end{align*}
    where we have used the fact that:
    \begin{align*}
        \sum_{a' \in \arcsMod_{i_a}^+} e^{-\beta \costToGo_{a'}} \frac{\partial \costToGo_{a'}}{\partial \costToGo_{\bar a}} &= \sum_{a' \in \arcsMod_{i_a}^+} e^{-\beta \costToGo_{a'}} \cdot \textbf{1}\{a' = \hat a\} \\
        &\leq \max_{a' \in \arcsMod_{i_a}^+} e^{-\beta \costToGo_{a'}}.
    \end{align*}
    Thus, applying the triangle inequality, we obtain:
    \begin{align*}
        \left| \frac{\partial \hat \rho_a}{\partial \costToGo_{\bar a}} \right| &= \beta + \beta = 2\beta.
    \end{align*}
    
    This concludes the proof for this part of the proposition.
    
    \item For each $a, a' \in \arcsMod$:
    {\small
    \begin{align*}
        &\frac{r_{[a]}}{\toll_{[a']}}(\toll) \\
        = \hspace{0.5mm} &- \frac{\partial \toll_{[a]}}{\partial \toll_{[a']}} + \left( \frac{d\latency_{[a]}}{d\arcLoadMod}\big(\bar \arcLoadMod_{[a]}^\beta(\toll) \big) + \bar \arcLoadMod_{[a]}^\beta(\toll) \frac{d^2 \latency_{[a]}}{d\arcLoadMod^2} \big(\bar \arcLoadMod_{[a]}^\beta(\toll) \big) \right) \\
        &\hspace{1cm} \cdot \frac{\partial \bar \arcLoadMod_{[a]}^\beta}{\partial \toll_{[a']}}(\toll).
    \end{align*}
    }
    Define:
    \begin{align*}
        C_{dds} &:= \max_{x \in [0, g_o]}\left\{ \frac{d^2 \latency_{[a]}}{d\arcLoadMod^2}(x) \right\}.
    \end{align*}
    Meanwhile, by the first part of this proposition, and the Cauchy-Schwarz inequality:
    \begin{align*}
        \left|\frac{\partial \bar \arcLoadMod_{[a]}^\beta}{\partial \toll_{[a']}}(\toll) \right| &= \left| e_{[a]}^\top \frac{d \bar \arcLoadMod^\beta}{d\toll}(\toll) e_{[a']} \right| \\
        &\leq \left\Vert \frac{d \bar \arcLoadMod^\beta}{d\toll}(\toll) \right\Vert_2 \\
        &\leq \Vert B \Vert_2^2 \Vert C \Vert_2^2 \cdot \frac{1}{C_F}.
    \end{align*}
    Thus, we have:
    \begin{align*}
        \left|\frac{r_{[a]}}{\toll_{[a']}}(\toll)\right| &\leq 1 + (C_{ds} + g_o C_{dds}) \cdot \Vert B \Vert_2^2 \Vert C \Vert_2^2 \cdot \frac{1}{C_F} \\
        &:= C_r.
    \end{align*}
    Thus, $C_r > 0$ uniformly upper bounds the partial derivatives of $r_{[a]}$ over all of its components $\toll_{[a']}$ and all arguments $\toll \in \R^{|\arcsOrig|}$. This establishes the Lipschitz continuity of each $r_{[a]}$, and thus concludes the proof.
\end{enumerate}
\end{proof}

\subsubsection{Proof of Theorem \ref{Thm: Convergence, w, p, discrete}}
\label{subsubsec: A2, Proof of Thm: Convergence, w, p, discrete}

We complete the proof of Theorem \ref{Thm: Convergence, w, p, discrete}, restated below: There exists some $\epsilon > 0$ such that, if $\Vert \tollDiscrete[0] - \bar \toll \Vert_2 \leq \epsilon$, then (a):
\begin{align*}
    &\limsup_{n \ra \infty} \E\big[ \Vert \probDist[n] - \bar \probDist^\beta(\bar \toll) \Vert_2^2 + \Vert \tollDiscrete[n] - \bar \toll \Vert_2^2 \big] \\
    = \hspace{0.5mm} &O\left(\mu + \frac{a}{\mu} \right),
\end{align*}
and (b) for each $\delta > 0$:
\begin{align*}
    &\limsup_{n \ra \infty} \Prob \big[ \Vert \probDist[n] - \bar \probDist^\beta(\bar \toll) \Vert_2^2 + \Vert \tollDiscrete[n] - \bar \toll \Vert_2^2 \geq \delta \big] \\
    = \hspace{0.5mm} &O\left(\frac{\mu}{\delta} + \frac{a}{\delta \mu} \right).
\end{align*}
The result follows by applying the global convergence of the continuous-time toll dynamics \ref{Eqn: General Network, p flow, continuous} under the affine latency assumption, as provided by Lemma \ref{Lemma: Convergence, p, continuous}. Theorem \ref{Thm: Convergence, w, p, discrete} now follows by applying the two-timescale stochastic approximation results in Borkar \cite{Borkar2008StochasticApproximation}, Chapters 2 and 9.

\end{document}